\documentclass[12pt]{article}
\usepackage{graphicx} 
\graphicspath{{./figs/}}
\usepackage{subcaption}
\usepackage{amsmath,amssymb,xcolor,physics}
\usepackage{url}
\usepackage{ulem}
\usepackage{hyperref}
\usepackage{soul}  
\usepackage{tikz}
\usepackage{pgfplots}
\usepackage{cite}
\usetikzlibrary{fadings}
\pgfplotsset{compat=1.18}

\allowdisplaybreaks[4]

\begin{document}

\begin{center}
{\Large New phases in QCD at finite temperature and chemical potential}\\
\vspace{3mm}

\end{center}
\vspace{0.1cm}
\vspace{0.1cm}
\begin{center}
Masanori Hanada,$^{\rm Q}$ Jack Holden,$^{\rm C}$ and Hiromasa Watanabe$\, ^{\rm D}$
\end{center}
\vspace{0.3cm}

\begin{center}
$^{\rm Q}$\, School of Mathematical Sciences, Queen Mary University of London\\
Mile End Road, London, E1 4NS, United Kingdom\\
\vspace{1mm}
$^{\rm C}$\, 
Yau Mathematical Sciences Center,
Tsinghua University, Beijing, 100084, China\\
\vspace{1mm}
$^{\rm D}$\, Department of Physics, and Research and Education Center for Natural Sciences\\
Keio
University, 223-8521, Kanagawa, Japan
\end{center}

\vspace{0.5cm}

\begin{center}
  {\bf Abstract}
\end{center}
Understanding the phases of quantum chromodynamics (QCD) at finite temperature and baryon density is crucial for describing matter in heavy-ion collisions and the interior of neutron stars. While the transition from the confined phase to the deconfined phase has been extensively studied at vanishing chemical potential, recent theoretical work suggests the existence of intermediate phases with partial deconfinement, where only a subset of the color degrees of freedom is deconfined. 
In this paper, we study the generalization of partial deconfinement in QCD in the Veneziano large-$N_{\rm c}$ limit ($N_{\rm f}/N_{\rm c}$ fixed) to finite baryon chemical potential. We find that the partially deconfined phase has finer structure, and hence that there are four phases in QCD at finite temperature and moderately large chemical potential: complete confinement, complete deconfinement, and two kinds of partial deconfinement.
A key ingredient is the refined understanding of the meaning of the `Gross-Witten-Wadia point', i.e., the opening of a gap in the distribution of Polyakov line phases: either string condensation or baryon condensation causes the GWW transition. 
As a by-product, we observe the emergence of the QCD critical point from the interplay between baryon condensation and partial deconfinement. While our approach is necessarily qualitative due to the fermion sign problem, it provides a unified theoretical framework for understanding the rich phase structure of dense QCD matter and offers new perspectives on the location and nature of the QCD critical point.

\newpage
\tableofcontents

\section{Introduction}
\begin{figure}[htbp]
	\centering
\includegraphics[width=0.75\textwidth]{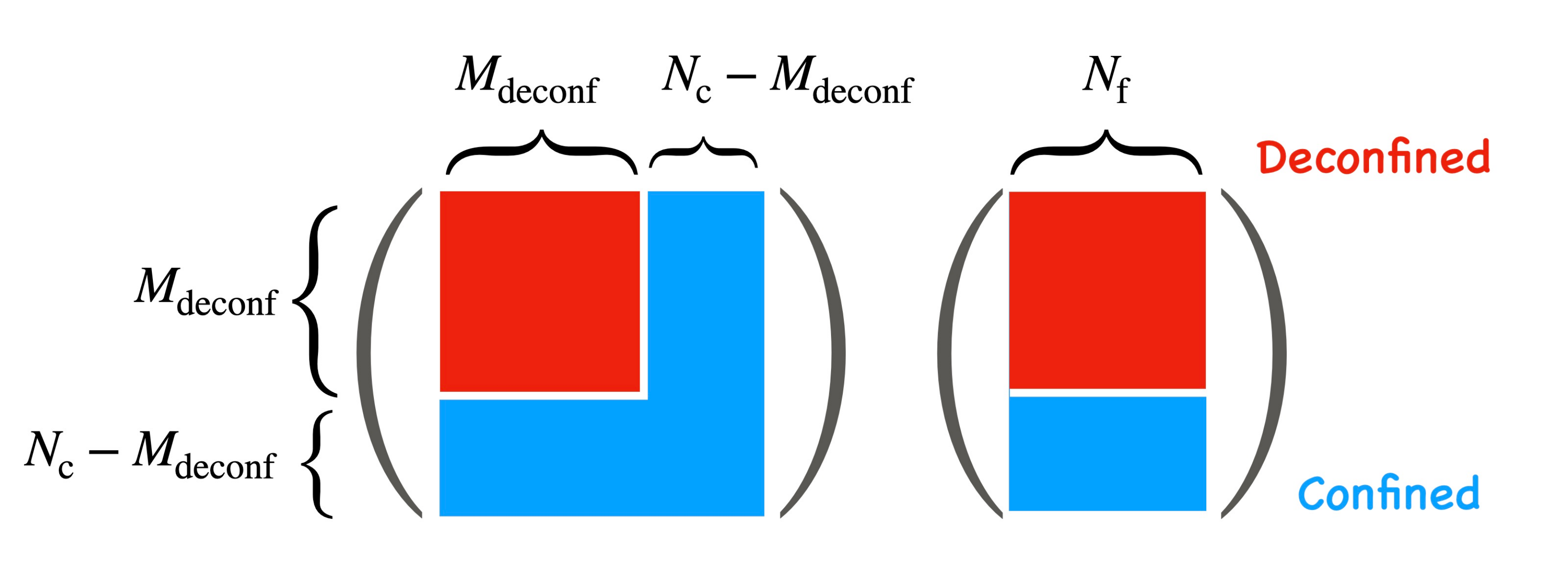}
	\caption{In the partially deconfined phase, an SU($M_{\rm deconf}$) subgroup of the SU($N_{\rm c}$) gauge group deconfines. $M_{\rm deconf}$ can grow from zero (complete confinement) to $N_{\rm c}$ (complete deconfinement). 
    }
	\label{fig:partial_deconf_in_components}
\end{figure}

\begin{figure}[htbp]
	\centering
\includegraphics[width=0.9\textwidth]{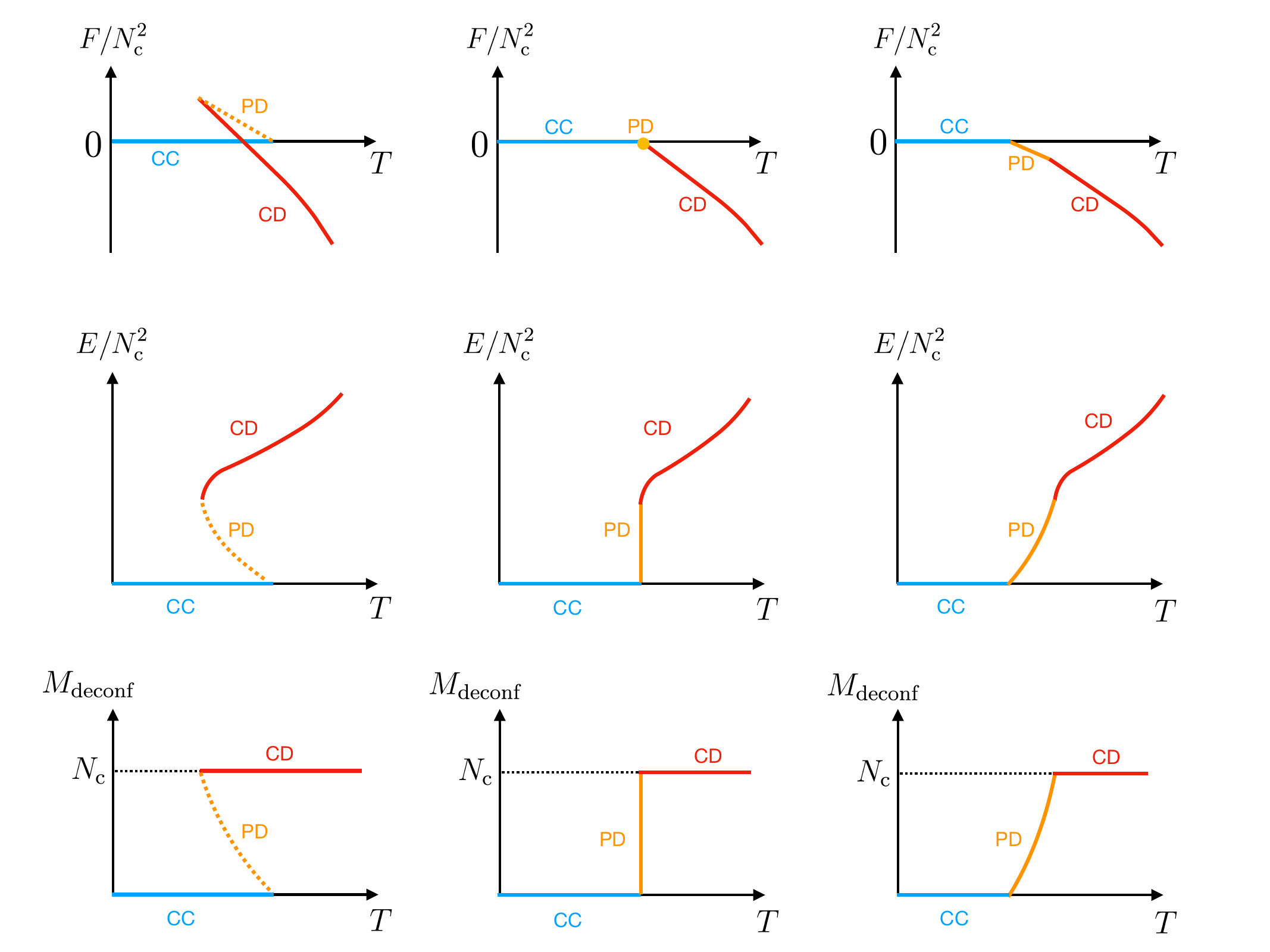}
	\caption{Three basic patterns of confinement/deconfinement transition at zero chemical potential. Free energy $F$, energy $E$, and the size of the deconfined sector $M_{\rm deconf}$ are shown. 
	\textcolor{blue}{Blue}, \textcolor{orange}{orange}, and \textcolor{red}{red} lines are \textcolor{blue}{Completely Confined (CC)}, \textcolor{orange}{Partially Deconfined (PD)}, and \textcolor{red}{Completely Deconfined (CD)} phases. The solid and dotted lines correspond to the minimum and maximum of the free energy, respectively.
    [Left] PD has a negative specific heat. Pure Yang-Mills theory, QCD in the 't Hooft large-$N_\mathrm{c}$ limit, and QCD in the heavy-quark region have such a phase structure. When the temperature is varied, a first-order transition with hysteresis is observed. 
    [Center] Free energy is degenerate on PD. 4D Yang-Mills theory compactified on a small sphere and the Gaussian matrix model exhibit such a phase structure. When the temperature is varied, a first-order transition without hysteresis is observed. 
    [Right] PD has positive specific heat. QCD in the Veneziano large-$N_\mathrm{c}$ limit with sufficiently large $N_{\rm f}/N_{\rm c}$ and light mass has such a phase structure. 
    }
\label{fig:three_patterns}
\end{figure}

Large-$N_\mathrm{c}$ gauge theories at finite temperature have multiple phases, not merely confinement and deconfinement. This became clear through the investigation of holographic duality that connects black holes to deconfinement~\cite{Witten:1998zw,Aharony:1999ti}: between the graviton gas and the large black hole, there is an intermediate, small black hole phase. 
Subsequent efforts clarified similar patterns more generically, without relying on the duality~\cite{Sundborg:1999ue,Aharony:2003sx}. Building on these, Partial Deconfinement (PD) was discovered~\cite{Hanada:2016pwv,Berenstein:2018lrm,Hanada:2018zxn} (Fig.~\ref{fig:partial_deconf_in_components} and Fig.~\ref{fig:three_patterns}). In the partially deconfined phase of an SU($N_{\rm c}$) gauge theory, an SU($M_{\rm deconf}$) subgroup deconfines, where $M_{\rm deconf}$ changes between $M_{\rm deconf}=0$ (Complete Confinement; CC) and $M_{\rm deconf}=N_{\rm c}$ (Complete Deconfinement; CD). Ref.~\cite{Hanada:2021ksu} discussed that partial deconfinement may be well-defined even at finite $N_{\rm c}$. An attempt to generalize partial deconfinement to SU(3) QCD has been made in refs.~\cite{Hanada:2023krw,Hanada:2023rlk}. 

An important feature of partial deconfinement is that it is valid even for theories without center symmetry, most notably, QCD. Furthermore, even in theories with center symmetry, both partially deconfined and completely deconfined phases lie in the center-broken region, and hence these phases cannot be distinguished based on the center symmetry. In fact, the `symmetry' behind partial deconfinement is gauge symmetry~\cite{Hanada:2020uvt}, as reviewed in Sec.~\ref{sec:PD-review}. The opening of a gap in the distribution of Polyakov line phases --- which we call the `Gross-Witten-Wadia (GWW) transition' --- plays an important role. Note that the term `GWW transition' is often used for a more specific case, i.e., the third-order transition in the GWW model~\cite{Gross:1980he,Wadia:2012fr}. We use the same word to refer more generally to the opening of the gap, regardless of other details such as the order of the phase transition. We hope that this is not considered an abuse of terminology, because the underlying physical mechanism is the same regardless of such details. 

\begin{figure}[htbp]
	\centering
\includegraphics[width=10cm]{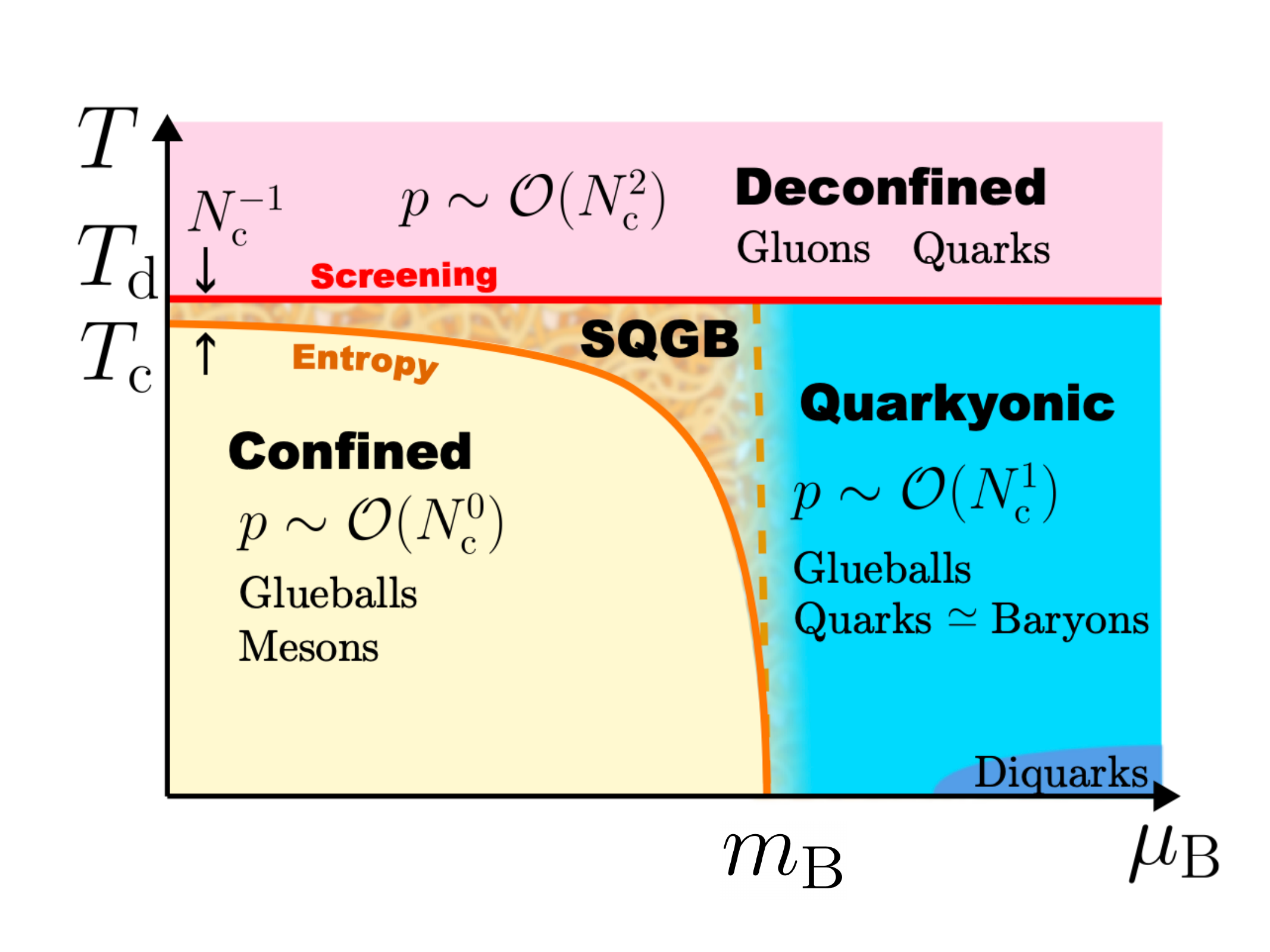}
\includegraphics[width=11cm]{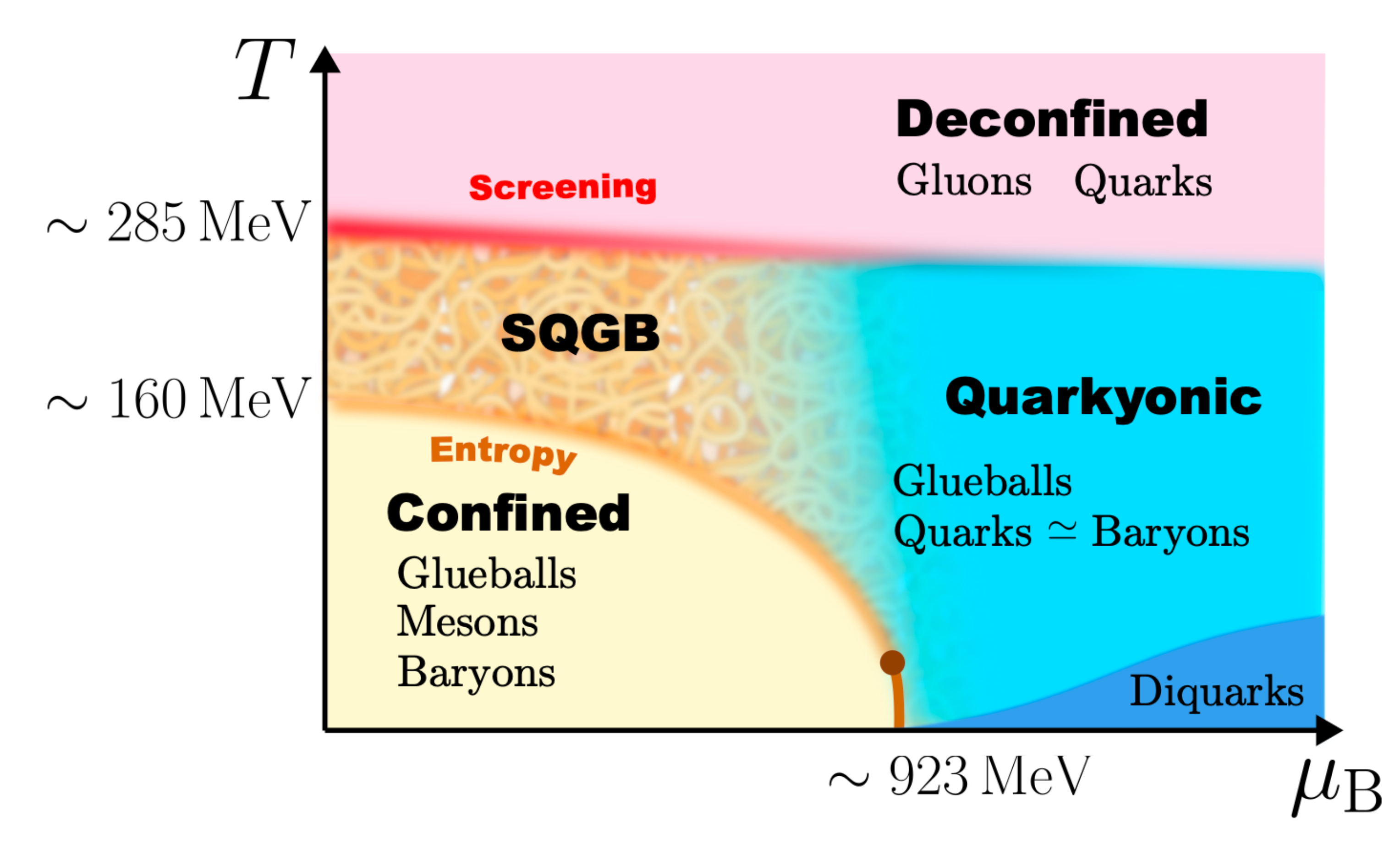}
	\caption{QCD phase diagram proposed by Fujimoto, Fukushima, Hidaka, and McLerran. [Top] 't Hooft large-$N_{\rm c}$ limit plus finite-$N_{\rm c}$ corrections. [Bottom] SU(3). These pictures are from ref.~\cite{Fujimoto:2025sxx}.
    Here, $m_{\rm B}$ is the baryon mass, and $\mu_{\rm B}$ and $\mu_{\rm q}$ are baryon and quark chemical potentials related by $\mu_{\rm B}=N_{\rm c}\mu_{\rm q}$. Note that $m_{\rm B}$ was denoted as $M_{\rm B}$ in ref.~\cite{Fujimoto:2025sxx}; we have changed the notation to avoid a potential confusion with $M_{\rm deconf}$ and $M_{\rm active}$. 
    }
\label{fig:FFHM_phase_diagram}
\end{figure}

Stimulated by partial deconfinement, there have been several attempts to establish the three phases phenomenologically; see e.g., refs.\cite{Glozman:2019fku,Cohen:2023hbq,Fujimoto:2025sxx}. Among them, Fujimoto, Fukushima, Hidaka, and McLerran (FFHM)~\cite{Fujimoto:2025sxx} proposed the Spaghetti of Quarks with Glueballs (SQGB) (Fig.~\ref{fig:FFHM_phase_diagram}). As we shall discuss in this paper, the SQGB can naturally be understood as partial deconfinement, although some modifications are needed. Still, ref.~\cite{Fujimoto:2025sxx} contains an interesting new ingredient: they considered QCD at finite temperature ($T$) and finite baryon chemical potential ($\mu_{\rm B}$), and discussed a potential connection between SQGB and the quarkyonic phase~\cite{McLerran:2007qj}. 
This gave us motivation to investigate partial deconfinement at finite baryon density. Although some speculation was made regarding the connection between partial deconfinement and the QCD critical point~\cite{Hanada:2018zxn}, no systematic investigation was undertaken.

To make clearer statements, it is useful to take the large-$N_\mathrm{c}$ limit. In much of the literature concerning QCD, including FFHM~\cite{Fujimoto:2025sxx}, the 't Hooft large-$N_\mathrm{c}$ limit (i.e., the number of colors $N_{\rm c}$ is sent to infinity while the number of flavors $N_{\rm f}$ is fixed) was considered. Although this limit is useful for high- or low-temperature regimes, it is not the ideal setup to study the entire phase diagram of real-world QCD, because even the order of the phase transition does not match. Specifically, although the partially deconfined phase is stable in real-world QCD, it turns into the unstable saddle in the canonical partition function in the 't Hooft large-$N_\mathrm{c}$ limit. The appropriate setup that resembles the real-world QCD is the Veneziano large-$N_\mathrm{c}$ limit~\cite{Veneziano:1976wm} (i.e., $\eta\equiv\frac{N_{\rm f}}{N_{\rm c}}$ fixed), in which a partially deconfined phase can be stable if $\eta$ is sufficiently large and the quark mass is sufficiently small. 

\begin{figure}[htbp]
	\centering
\includegraphics[width=16cm]{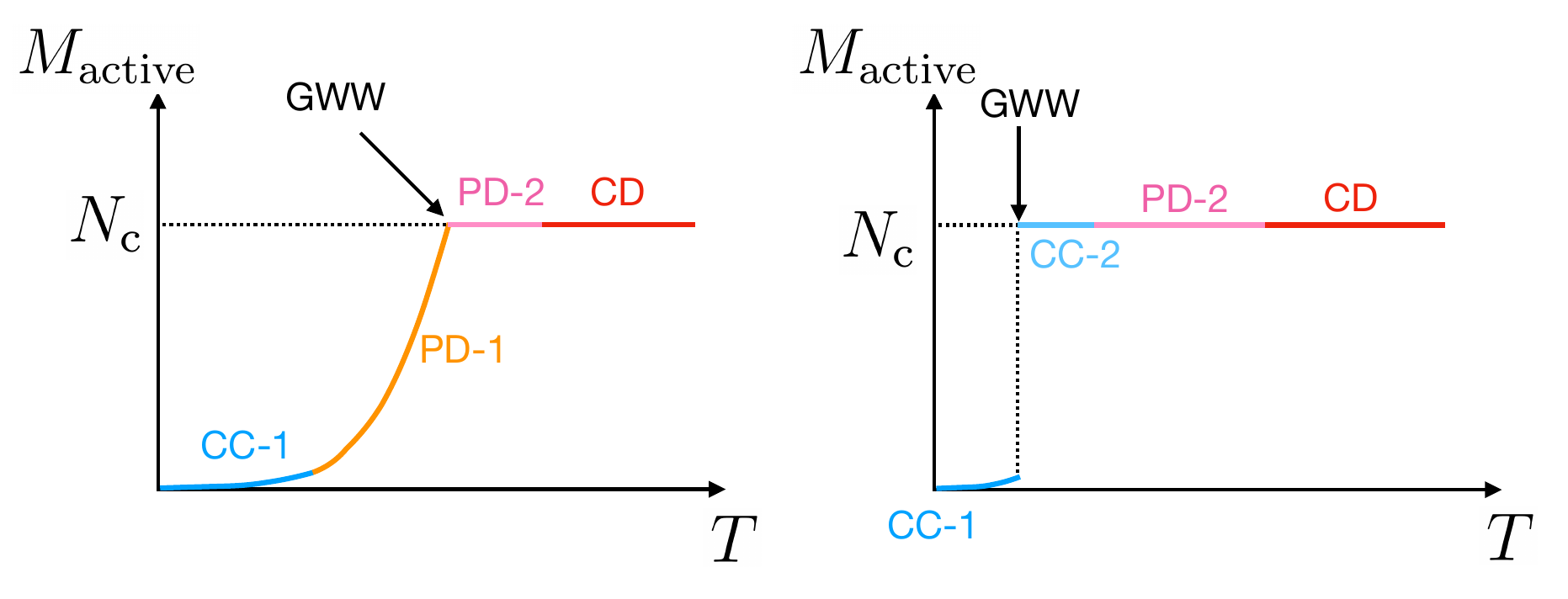}
	\caption{The revised picture of partial deconfinement in finite-density QCD in the Veneziano limit. The vertical axis $M_{\rm active}$ is color degrees of freedom, which are in either the deconfined sector (string-condensed sector) or the baryon-condensed sector. 
	[Left] Low-$\mu_{\rm q}$ regime. 
	The pink line indicates the new phase. `Complete confinement' (blue line; \textcolor{blue}{CC-1}) is likely to be an approximate notion and it could be smoothly connected to `partial deconfinement' (orange line; \textcolor{orange}{PD-1}). For this reason, in ref.~\cite{Hanada:2019kue}, they were not distinguished and both are called `partial deconfinement'. 
    The new phase depicted in pink (\textcolor{magenta}{PD-2}) is still partially deconfined. 
    \textcolor{blue}{CC-1} and \textcolor{orange}{PD-1} could be separated by a crossover rather than a phase transition.
	[Right] Large-$\mu_{\rm q}$ regime. Baryon condensation takes place before partial deconfinement sets in; we used \textcolor{blue}{CC-1} and \textcolor{cyan}{CC-2} to denote two `completely confined' phases (`confined' in the sense that long strings are not condensed). At higher temperatures, partial deconfinement (pink line; \textcolor{magenta}{PD-2}) sets in with the condensation of long strings. 
	We did not draw a line connecting \textcolor{blue}{CC-1} and \textcolor{cyan}{CC-2} because $M_{\rm active}$ does not change continuously. For example, a parametrically small amount of baryon condensation can already make $M_{\rm active}$ jump from $\sim 0$ to $N_{\rm c}$.  As discussed in Sec.~\ref{sec:quarkyonic}, \textcolor{cyan}{CC-2} could be identified with the quarkyonic phase. 
    \textcolor{cyan}{CC-2} and \textcolor{magenta}{PD-2} could be separated by a crossover rather than a phase transition.}\label{fig:Mactive-vs-T-QCD}
\end{figure}

\begin{figure}[htbp]
	\centering
\includegraphics[width=8cm]{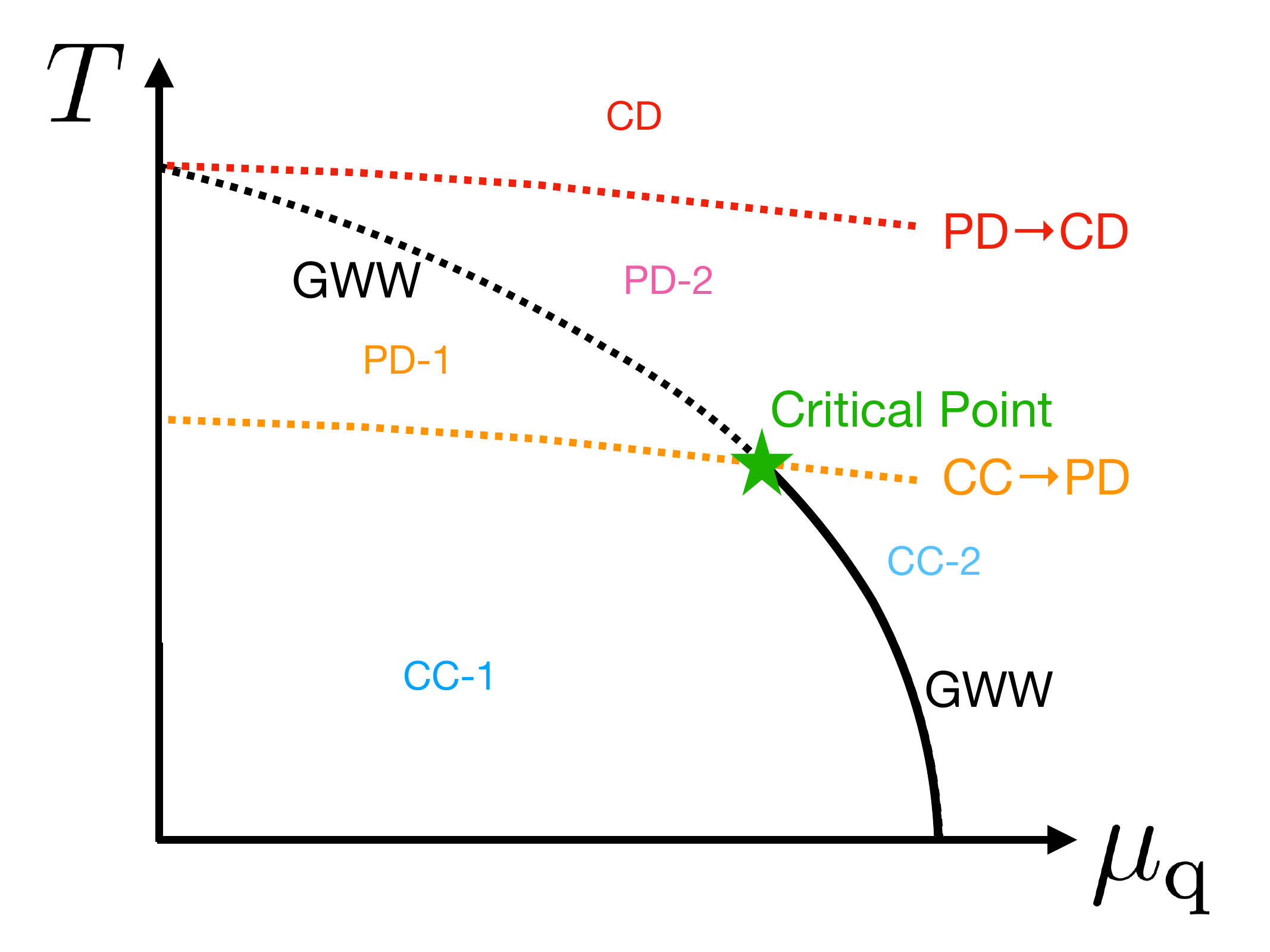}
	\caption{ Locations of each phase in the $T$-$\mu_{\rm q}$ plane. Three curves represent transition temperatures. The solid line represents a first-order transition. Note that CC$\to$PD is likely to be a crossover, and the `transition line' is only qualitative. 
    }\label{fig:revised_picture_mu_vs_T}
\end{figure}

In this paper, we study the Veneziano limit of QCD, with degenerate quark mass $m$. As a disclaimer, we cannot solve strongly-coupled physics directly at finite chemical potential. This is the case not just analytically but also numerically because of the fermion sign problem~\cite{deForcrand:2009zkb,Nagata:2021ugx}.\footnote{
The situation may change when a fault-tolerant digital quantum computer becomes available.
} Therefore, our analysis at finite chemical potential will be based on a few general principles and will necessarily be qualitative. That said, we use a key observation that does not depend on the details of the strongly-coupled dynamics: \uline{\textit{either string condensation or baryon condensation causes the GWW transition}}. (Here, `string condensation' refers to the condensation of long strings, whose lengths increase parametrically with $N_{\rm c}$.) Based on this, we propose a phase structure depicted in Fig.~\ref{fig:Mactive-vs-T-QCD} and Fig.~\ref{fig:revised_picture_mu_vs_T}. Specifically, at moderately large baryon chemical potential, we find three transition temperatures: $T_{{\rm PD}\to{\rm CD}}$, which separates partial deconfinement (PD) and complete deconfinement (CD); $T_{{\rm CC}\to{\rm PD}}$, which separates complete confinement (CC) and partial deconfinement;\footnote{Strictly speaking, CC$\to$PD should be seen as a crossover rather than a phase transition, as we will see in Sec.~\ref{sec:Veneziano-zero-mu}.
} and $T_{\rm GWW}$ where the GWW transition takes place. The GWW transition separates the partially deconfined phase into two, which we call PD-1 ($T_{{\rm CC}\to{\rm PD}}<T<T_{{\rm GWW}}$) and PD-2 ($T_{{\rm GWW}}<T<T_{{\rm PD}\to{\rm CD}}$). At zero chemical potential, the GWW transition and PD$\to$CD transition merge, PD-2 disappears, and the three-phase picture proposed in the past is recovered. At a larger chemical potential, there are two confined phases, CC-1 and CC-2, with and without baryon condensation. 

This paper is organized as follows. In the rest of this introductory section, we provide a short summary of our proposal. A short review of partial deconfinement at vanishing chemical potential is provided in Sec.~\ref{sec:PD-review}. Sec.~\ref{sec:Veneziano-zero-mu} discusses partial deconfinement in QCD in the Veneziano limit. Sec.~\ref{sec:Veneziano-nonzero-mu} is the main part of this paper and studies the phase structure of QCD in the Veneziano limit at nonzero chemical potential. We conclude this paper with some discussion in Sec.~\ref{sec:discussion}. 
A summary of notation is provided in Appendix~\ref{sec:notation}. 

\subsection{A summary of the proposal}\label{sec:proposal_short_summary}
This subsection provides a summary of our proposal for the phase diagram of QCD in the Veneziano large-$N_\mathrm{c}$ limit. The details will be provided in later sections. 
\subsubsection{Vanishing baryon chemical potential}
To make this paper self-contained, we start with the case of zero chemical potential, which has been studied in the past. In this case, there are three phases: CC, PD, and CD.\\

\noindent
\textbf{Complete Confinement (CC):} At low temperatures $T < T_{\text{CC}\to\text{PD}}$, all color degrees of freedom remain confined. The Polyakov loop eigenvalue distribution is approximately uniform, reflecting the gauge-invariant nature of the ground state. Long strings (both open and closed) are not condensed, and the system exhibits the familiar properties of the hadronic phase.\\

\noindent
\textbf{Partial Deconfinement (PD):} For intermediate temperatures $T_{\text{CC}\to\text{PD}} < T < T_{\text{PD}\to\text{CD}}$, an $\textrm{SU}(M_{\rm deconf})$ subgroup of the $\textrm{SU}(N_{\rm c})$ gauge symmetry deconfines while the remaining $\textrm{SU}(N_{\rm c}-M_{\rm deconf})$ sector stays confined. This phase is characterized by:
\begin{itemize}
\item A non-uniform, ungapped Polyakov loop eigenvalue distribution
\item Condensation of strings localized in the $\textrm{SU}(M_{\rm deconf})$ sector
\item Gradual restoration of chiral symmetry as $M_{\rm deconf}$ increases
\item Coexistence of confined glueball-like and hadron-like states and deconfined quark-gluon degrees of freedom
\end{itemize}

\noindent
\textbf{Complete Deconfinement (CD):} At high temperatures $T > T_{{\rm PD}\to{\rm CD}}$, all colors deconfine ($M_{\rm deconf}=N_{\rm c}$). The Polyakov loop eigenvalue distribution develops a gap, signaling the Gross-Witten-Wadia (GWW) transition. This corresponds to the familiar quark-gluon plasma phase.\\

The transition temperatures are estimated as $T_{\text{CC}\to\text{PD}} \sim 175$ MeV and $T_{\text{PD}\to\text{CD}} \sim 350$ MeV based on lattice QCD analysis~\cite{Hanada:2023krw,Hanada:2023rlk} (Table~\ref{table:transition_temperature_estimate}). At zero chemical potential, the GWW transition coincides with the PD$\to$CD transition: $T_{\text{GWW}} = T_{\text{PD}\to\text{CD}}$.

\subsubsection{Finite baryon chemical potential}
At finite baryon chemical potential, the phase structure becomes even richer, leading to our central proposal of four distinct phases in QCD at moderately large chemical potential.\\

\begin{figure}[htbp]
	\centering
\includegraphics[width=16cm]{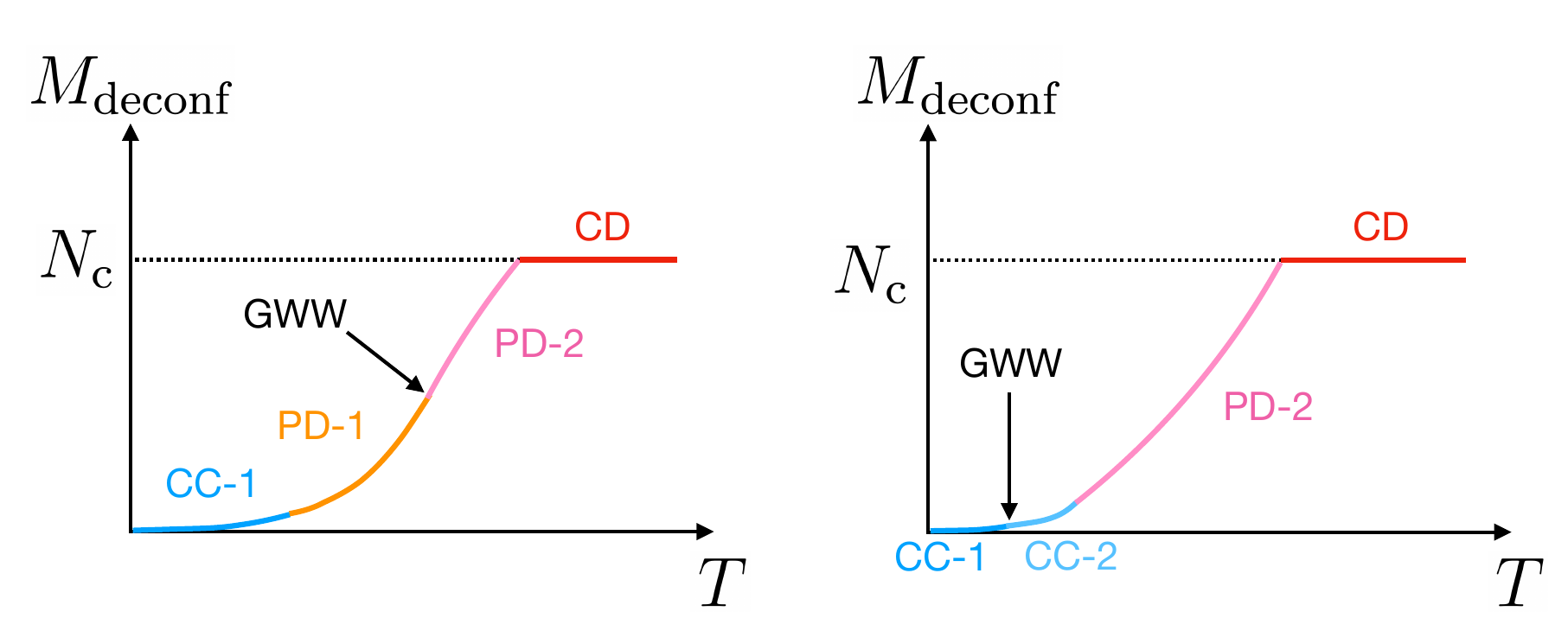}
	\caption{Conjectured behavior of the size of the deconfined sector $M_{\rm deconf}$ at finite chemical potential, which is different from $M_{\rm active}$ shown in Fig.~\ref{fig:Mactive-vs-T-QCD} (color degrees of freedom which are in either the deconfined sector or the baryon-condensed sector).
    }
	\label{fig:Mdeconf-vs-T-QCD}
\end{figure}

\noindent
\textbf{Two Partially Deconfined Phases (PD-1 and PD-2):} 
At nonzero but not too large values of chemical potential, there are two partially deconfined phases; see Fig.~\ref{fig:Mactive-vs-T-QCD}, Fig.~\ref{fig:revised_picture_mu_vs_T}, and Fig.~\ref{fig:Mdeconf-vs-T-QCD}.
Below the GWW transition temperature ($T < T_{\text{GWW}}$), partial deconfinement occurs while maintaining some confined color degrees of freedom. The Polyakov loop distribution remains ungapped. We call this phase PD-1. 
Above the GWW transition temperature ($T > T_{\text{GWW}}$), the system exhibits both baryon condensation and partial deconfinement. The Polyakov loop distribution is gapped, but deconfinement affects only $M_{\text{deconf}} < N_{\rm c}$ colors.
We call this phase PD-2. \\

\noindent
\textbf{Two Completely Confined Phases (CC-1 and CC-2):} At large values of chemical potential, there are two completely confined phases; see Fig.~\ref{fig:Mactive-vs-T-QCD}, Fig.~\ref{fig:revised_picture_mu_vs_T}, and Fig.~\ref{fig:Mdeconf-vs-T-QCD}.
At low temperatures, the system remains in the traditional confined phase with no baryon condensation and no deconfinement. We call this phase CC-1. 
As temperature increases, baryons condense at temperature $T_{\rm B}$ before deconfinement occurs. This baryon condensation creates non-trivial gauge orbits in the extended Hilbert space, leading to gap formation in the Polyakov loop eigenvalue distribution. Crucially, this represents a GWW transition driven by baryon condensation rather than string condensation.

\subsubsection{Connection to Existing Proposals}
Our framework provides a unified understanding of several existing ideas:
\begin{itemize}
\item The ``Spaghetti of Quarks with Glueballs" phase~\cite{Fujimoto:2025sxx} corresponds to partially deconfined phases. 

\item The CC-2 phase connects to the quarkyonic phase~\cite{McLerran:2007qj} in the 't Hooft limit. 

\item The gradual chiral symmetry restoration occurs throughout the partially deconfined phases. 

\item The QCD critical point emerges naturally from the interplay between baryon condensation and partial deconfinement. 
\end{itemize}

\section{Partial deconfinement at vanishing chemical potential: a short review}\label{sec:PD-review}

The partially deconfined phase is a phase of matter intermediate between the confined and deconfined phases. Previous work on partial deconfinement has focused on the special case of zero chemical potential. We will now review this special case. This section concerns generic gauge theories, without assumptions on spacetime dimensions or matter contents. 

Heuristically, one can imagine that some of the color degrees of freedom are in the confined phase, while the others are in the deconfined phase. This is analogous to familiar situations in which two phases occupy different spatial regions, such as ice and liquid water at the transition temperature --- but now the separation is taking place in the internal space (color space). 

There are significant differences from the familiar case of spatial separation. First, the interactions in different regions of color space are nonlocal. Second, due to the gauge symmetry, there is a gauge enhancement effect for states with `unbroken gauge symmetry'. (This term should not confuse the reader who will be aware that `gauge symmetry' is a redundancy in our description. We explain this in the next subsection, Sec.~\ref{sec:underlying_mechanism}.). At large $N_{\rm c}$ and zero chemical potential, the transition between the partially and completely deconfined phases is associated with the Polyakov loop eigenvalues undergoing a Gross-Witten-Wadia (GWW) transition. 

For QCD in the Veneziano large-$N_\mathrm{c}$ limit with sufficiently large $\eta=N_{\rm f}/N_{\rm c}$ and light quark mass, the crossover region at zero chemical potential corresponds to the partially deconfined phase, connecting complete confinement to complete deconfinement. In this case, the phase diagram for the partially deconfined phase follows the third column of Fig.~\ref{fig:three_patterns}, mirroring the crossover.

A finite $N_{\rm c}$ generalization of the GWW transition is the condensation of higher-representation Polyakov loops, so this has been proposed to characterize the partially deconfined phase in QCD at finite $N_{\rm c}$. It has also been conjectured that instanton condensation can mark the same transition point. Evidence supporting both conjectures has been given in \cite{Hanada:2023rlk, Hanada:2023krw}.  
Moreover, a reasonable prediction is that chiral symmetry breaking also coincides with this transition point. One argument follows from 't Hooft anomaly matching, together with the conceptual picture that the color sector contains both confined and deconfined degrees of freedom, implying that both chiral and center symmetry must be broken in the partially deconfined phase. The association between instanton condensation and chiral symmetry breaking lends further support to this idea. Finally, since chiral symmetry breaking by itself implies a transition, aligning it with the partial deconfinement transition avoids the need to introduce an additional phase.

\subsection{Polyakov loop and underlying mechanism of partial deconfinement}\label{sec:underlying_mechanism}
It is well known that the Polyakov loop in the fundamental representation is an order parameter for confinement in theories without explicitly broken center symmetry. It corresponds to the expectation value for the insertion of an isolated quark. 

\begin{figure}[htbp]
	\centering
\includegraphics[width=8cm]{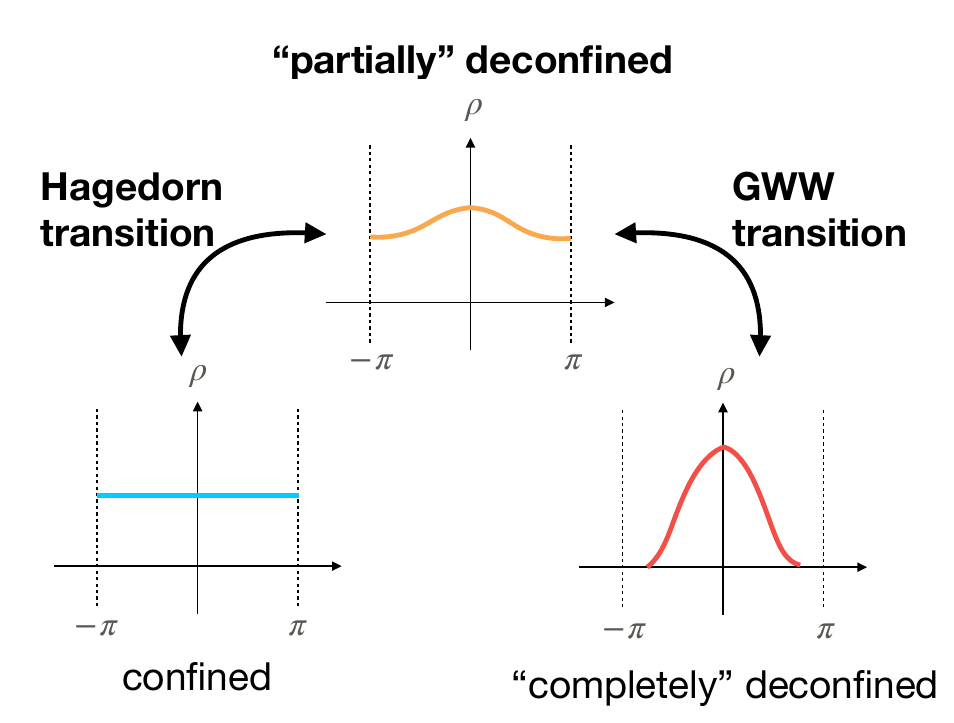}
	\caption{
    The relationship between the Polyakov line phase distribution and three phases at zero chemical potential. This figure is taken from ref.~\cite{Hanada:2019rzv}. 
    }\label{fig:pol_phase_dist}
\end{figure}

The eigenvalues of the Polyakov loop themselves represent a more refined gauge-invariant measurement, being gauge-invariant up to permutation of the eigenvalues. This has proven to be a useful tool to describe the partially deconfined phase. We can describe the Polyakov loop $P$ in terms of the eigenvalues $e^{\mathrm{i}\theta_j}$ of the holonomy (Polyakov line) $\mathcal{P}$ by
\begin{equation}
    P 
    \equiv
    \frac{1}{N_{\rm c}}\mathrm{Tr}
    \mathcal{P} 
    \equiv
    \frac{1}{N_{\rm c}} \sum_{j=1}^{N_{\rm c}} e^{\mathrm{i}\theta_j}\, ,  
\label{eq:Polyakov-loop}
\end{equation}
\begin{equation}
    \mathcal{P} 
    \equiv
    \mathrm{Path}\left( \exp\left(\mathrm{i}\int_0^\beta \mathrm{d} t \,A_t\right)\right)\, , 
\label{eq:Polyakov-line}
\end{equation}
where ``Path" denotes the path ordering.
At large $N_{\rm c}$, we can treat the distribution of eigenvalues as a continuum and introduce the eigenvalue density, $\rho(\theta)$, as in,
\begin{align}
    \frac{1}{N_{\rm c}} \sum_{j=1}^N f(\theta_j)
    =
    \int_{-\pi}^{+\pi} \mathrm{d}\theta\; \rho(\theta)\, f(\theta)\, 
\end{align}
for any function $f(\theta)$. 
This eigenvalue density offers a sharp characterization of partial deconfinement at zero chemical potential without relying on the center symmetry; see Fig.~\ref{fig:pol_phase_dist}. We can separate the $N_{\rm c}$ colors into $M_{\rm deconf}$ deconfined colors and $N_{\rm c}-M_{\rm deconf}$ confined colors. The confined sector of the color degrees of freedom can be represented as a uniformly-distributed contribution to the eigenvalue density distribution. The deconfined sector lacks any uniform component, and must have a `gap' of zero support in some neighborhood of $\theta=\pm \pi$. Combining these, we are left with a distribution that is \emph{non-uniform ungapped}. This signifies the partially deconfined phase, and gives a precise interpretation of the cartoon in Fig.~\ref{fig:partial_deconf_in_components}.

The ungapped non-uniform distribution then represents a partially deconfined phase. The height of the uniform sector corresponds to the $N_{\rm c}-M_{\rm deconf}$ confined colors. The critical point between the gapped and ungapped non-uniform distributions corresponds to the famous GWW transition. This indicates the transition between the completely deconfined and the partially deconfined phases.

The above statements are not obvious and require a derivation. As a simple example, let us consider the SU($N_{\rm c}$)-gauged Gaussian matrix model. The dynamical degrees of freedom are $N_{\rm c}\times N_{\rm c}$ Hermitian matrices $X_1,\cdots,X_d$, where $d\ge 2$ is the number of matrices. The Hamiltonian is given by 
\begin{align}
\hat{H}
=
{\rm Tr}
\sum_{I=1}^d\left(
\frac{1}{2}\hat{P}_I^2
+
\frac{1}{2}\hat{X}_I^2
\right)\, .  
\end{align}
This Hamiltonian is invariant under the adjoint action of SU($N_{\rm c}$) defined by
\begin{align}
\hat{X}_{I,ij}
\to
(g\hat{X}_{I}g^{-1})_{ij}
=
\sum_{k,l}g_{ik}\hat{X}_{I,kl}g^{-1}_{lj}, 
\qquad
\hat{P}_{I,ij}
\to
(g\hat{P}_{I}g^{-1})_{ij}
=
\sum_{k,l}g_{ik}\hat{P}_{I,kl}g^{-1}_{lj}\, . 
\end{align}
This SU($N_{\rm c}$) symmetry is treated as the gauge symmetry. This model exhibits the confinement/deconfinement transition. The deconfinement is caused by string condensation, as we will see shortly.  

One way of gauging SU($N_{\rm c}$) is to restrict the Hilbert space to SU($N_{\rm c}$)-invariant states. Such states form the gauge-invariant Hilbert space ${\cal H}_{\rm inv}$. By using ${\cal H}_{\rm inv}$, the canonical partition function at temperature $T$ can be written as 
\begin{align}
Z(T)
&=
{\rm Tr}_{\mathcal{H}_{\rm inv}}\left(
e^{-\hat{H}/T}
\right)\, . 
\label{eq:canonical_Z-H-inv}
\end{align}
Equivalently, we can select the extended Hilbert space ${\cal H}_{\rm ext}$ that contains states not invariant under SU($N_{\rm c}$). 
As a basis for ${\cal H}_{\rm ext}$, we can use the coordinate eigenstates  $\ket{X}$ that satisfy $\hat{X}_{I,ij}\ket{X}=X_{I,ij}\ket{X}$:
\begin{align}
\mathcal{H}_{\rm ext}
=
\textrm{Span}\{
\ket{X}|X\in\mathbb{R}^{dN^2}
\}. 
\end{align}
We can also use the momentum eigenstates $\ket{P}$ that satisfy $\hat{P}_{I,ij}\ket{P}=P_{I,ij}\ket{P}$:
\begin{align}
\mathcal{H}_{\rm ext}
=
\textrm{Span}\{
\ket{P}|P\in\mathbb{R}^{dN^2}
\}. 
\end{align}
The SU($N_{\rm c}$) gauge transformation acts on these states as 
\begin{align}
\ket{X}
\to
\ket{g^{-1} Xg}\, , 
\qquad
\ket{P}
\to
\ket{g^{-1} Pg}\, . 
\end{align}
In the extended Hilbert space, the states related by an SU($N_{\rm c}$) transformation should be identified. 
The canonical partition function \eqref{eq:canonical_Z-H-inv} can also be written as
\begin{align}
Z(T)
&=
\frac{1}{{\rm vol}G}\int_G \mathrm{d} g\,
{\rm Tr}_{{\cal H}_{\rm ext}}\left(
\hat{g}
e^{-\hat{H}/T}
\right)\, . 
\label{eq:canonical_Z-H-ext}
\end{align}
$G=\textrm{SU}(N_{\rm c})$ denotes the gauge group, with $g$ representing an element of the group and $\hat{g}$ its action on the extended Hilbert space ${\cal H}_{\rm ext}$ (in this context, through the adjoint representation). The integration is performed using the Haar measure.

Readers may recognize an analogous construction in the context of discrete gauge groups. There, the insertion of $\hat{g}$ corresponds to imposing a twisted boundary condition, and gauging the discrete symmetry involves summing over all such twists.

The expression \eqref{eq:canonical_Z-H-ext} is directly related to the path integral with temporal gauge field $A_t$. 
The group element $g$ corresponds to the Polyakov line $\mathcal{P}$ in the Euclidean path integral~\cite{Hanada:2020uvt}. 

From a state $|\phi\rangle\in\mathcal{H}_{\rm ext}$, we can obtain a singlet $\hat{\pi}|\phi\rangle\in\mathcal{H}_{\rm inv}$, where $\hat{\pi}\equiv\frac{1}{{\rm vol}G}\int_Gdg\hat{g}$. 
With this correspondence, we can describe the same physics by using $\mathcal{H}_{\rm ext}$ or $\mathcal{H}_{\rm inv}$. 

Let us use the Fock basis for $\mathcal{H}_{\rm ext}$, which provides us with the energy eigenstates. 
A specific Fock state $\ket{\phi}$ contributes to the canonical partition function as
\begin{align}
e^{-E_\phi/T}\cdot
\frac{1}{{\rm vol}G}\int_G \mathrm{d} g
\bra{\phi}
\hat{g}
\ket{\phi}\, .  
\end{align}
For generic high-energy states, $\bra{\phi}\hat{g}\ket{\phi}$ quickly decays as $g$ departs from the identity. On the other hand, if the excitations are restricted in the SU($M_{\rm deconf}$) sector as in Fig.~\ref{fig:partial_deconf_in_components}, there is a large stabilizer subgroup SU($N_{\rm c}-M_{\rm deconf}$) that leaves $\ket{\phi}$ unchanged (i.e., $\hat{g}\ket{\phi}=\ket{\phi}$), and a large enhancement factor proportional to the volume of SU($N_{\rm c}-M_{\rm deconf}$), which behaves as an exponential of $(N_{\rm c}-M_{\rm deconf})^2$, arises. This enhancement factor triggers the partial deconfinement. 

The ground state has the largest stabilizer --- SU($N_{\rm c}$) itself --- and the orbit is just a single point. Any group element $g$ contributes equally. Because $g$ is the same as the Polyakov line, the distribution of the Polyakov line in the path integral becomes Haar random, leading to the uniform phase distribution. 

For generic SU($M_{\rm deconf}$)-deconfined states, the stabilizer is SU($N_{\rm c}-M_{\rm deconf}$). Up to the choice of the embedding of SU($M_{\rm deconf}$) into SU($N_{\rm c}$), the Polyakov line can take the block diagonal form $\begin{pmatrix}\mathcal{P}_{\rm dec} & 0\\
0 & \mathcal{P}_{\rm con}\end{pmatrix}$, where $\mathcal{P}_{\rm con} \in \mathrm{SU}(N_{\rm c}-M_{\rm deconf})$ is Haar random and leads to a constant offset in the phase distribution. 

This mechanism works for many theories, including Yang-Mills theory and QCD. In Sec.~\ref{sec:Veneziano-zero-mu}, we discuss partial deconfinement in QCD in the Veneziano large-$N_{\rm c}$ limit and at zero chemical potential. 

In Sec.~\ref{sec:Veneziano-nonzero-mu}, we will introduce a chemical potential. A new phenomenon at finite chemical potential is that baryon condensation leads to a nontrivial gauge orbit. Therefore, we will have to modify the argument above that assumed deconfinement (condensation of strings) as the only source of nontrivial gauge orbits. What we can see from the distribution of Polyakov line phases is the number of `active' degrees of freedom. Specifically, $M_{\rm active}$ will be the number of degrees of freedom either in the deconfined sector or in the baryon condensation sector. (Below, we interpret ``deconfinement" as the condensation of strings, and hence, baryon-condensed states without string condensation are in the confined phase.) For this purpose, it is useful to see partial deconfinement in terms of operators. In the case of the Gaussian matrix model, gauge-invariant states can be obtained by acting with string creation operators $\mathrm{Tr}(\hat{A}^\dagger_{I_1}\hat{A}^\dagger_{I_2}\cdots)$, where $\hat{A}^\dagger_I = \frac{\hat{X}_I-\mathrm{i}\hat{P}_I}{\sqrt{2}}$, on the Fock vacuum. High-energy, deconfined states are dominated by long strings (long traces) whose length is of order $N_{\rm c}^2$ --- hence the deconfinement is string condensation (condensation of long strings whose lengths are of order $N_{\rm c}^2$). Up to the zero-point contribution ($E_0=\frac{dN_{\rm c}^2}{2}$), the energy is proportional to the total length of the strings (number of creation operators $\hat{A}^\dagger$). When the energy is $E=\frac{M_{\rm deconf}^2}{4}+E_0$ ($M_{\rm deconf}<N_{\rm c}$), SU($M_{\rm deconf}$)-deconfinement is realized, i.e., the excitations are localized in the SU($M_{\rm deconf}$) sector up to gauge transformation. Therefore, most of the string operators with length $L=\frac{M_{\rm deconf}^2}{4}$ are obtained by symmetrizing the string operators localized in the SU($M_{\rm deconf}$)-sector depicted in Fig.~\ref{fig:partial_deconf_in_components}. If such operators act on states with nontrivial gauge orbit such as baryon-condensed states, strings are localized in the SU($M_{\rm deconf}$) sector, and hence, such states are understood as partially deconfined states. The same argument is applicable to other theories, including QCD. We will see such a situation in Sec.~\ref{sec:residual_symmetry_after_baryon_condensation}. 

\begin{figure}[htbp]
	\centering
\includegraphics[width=0.4\textwidth]{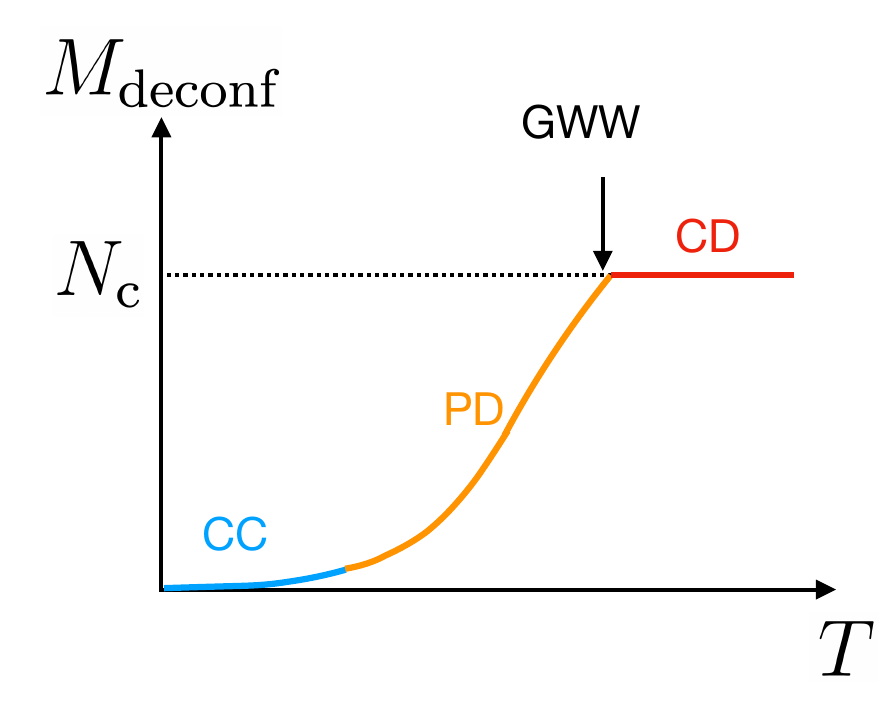}
	\caption{Weakly coupled QCD. 
    Strictly speaking, complete confinement (\textcolor{blue}{CC}) in the literal sense is achieved only at $T=0$. However, $M_{\rm deconf}/N_{\rm c}$ is parametrically suppressed at low temperature, and hence, the low-temperature regime can be seen as effectively completely confined. \textcolor{blue}{CC} and \textcolor{orange}{PD} are not separated by a sharp phase transition, while \textcolor{orange}{PD} and \textcolor{red}{CD} are separated by a third-order GWW transition. 
    }\label{fig:partial_deconf_QCD}
\end{figure}

\section{QCD in the Veneziano limit at vanishing chemical potential}\label{sec:Veneziano-zero-mu}
In this section, we summarize the properties of the partially deconfined phase in QCD in the Veneziano large-$N_\mathrm{c}$ limit. In addition, we explain how the proposed features of the SQGB phase follow from partial deconfinement. Partial deconfinement provides us with a more refined picture and addresses several issues with the SQGB proposal. 

In Sec.~\ref{sec:Veneziano-nonzero-mu}, we will discuss that the partially deconfined phase splits into two phases --- PD-1 and PD-2 --- when the baryon chemical potential is turned on. At zero chemical potential, we expect only PD-1, and hence we will discuss the similarities and differences between PD-1 and SQGB. 
\subsection{Partial deconfinement in the Veneziano limit}
Weak coupling (small volume) analysis of QCD on a three-sphere in the Veneziano limit~\cite{Hanada:2019kue}, based on computations in ref.~\cite{Schnitzer:2004qt}, established partial deconfinement as schematically depicted in Fig.~\ref{fig:partial_deconf_in_components} and Fig.~\ref{fig:partial_deconf_QCD}. 
While complete confinement takes place only at $T=0$ in the literal sense, the deconfined sector is parametrically small at low temperature as $\sim e^{-1/(R_{\mathrm{S}^3}T)}$ where $R_{\mathrm{S}^3}$ is the radius of the three-sphere, and hence, practically, the contribution from the deconfined sector is visible only at moderately high temperatures, $T\gtrsim T_{{\rm CC}\to{\rm PD}}\sim 1/R_{\mathrm{S}^3}$. 

The general mechanism of partial deconfinement~\cite{Hanada:2020uvt} (Sec.~\ref{sec:underlying_mechanism}) is based only on symmetry. As long as the distribution of Polyakov line phases changes continuously, the same picture is valid at strong coupling (large volume) as well.\footnote{
The Polyakov line considered here is defined at each spatial point. The phases are evaluated before taking the spatial average.}$^,$ 
\footnote{Note, however, that this picture needs to be improved at finite baryon chemical potential. Indeed, in Sec.~\ref{sec:Veneziano-nonzero-mu}, we will find that another mechanism different from partial deconfinement can also lead to the formation of a gap in the Polyakov line phase distribution. 
} A difference is that, at strong coupling, in the 't Hooft limit ($\eta=0$), the deconfinement transition is of first order with hysteresis (the first column in Fig.~\ref{fig:three_patterns}). For the thermal phase transition to be not of first order as in real-world QCD, $\eta\equiv\frac{N_{\rm f}}{N_{\rm c}}$ needs to be sufficiently large and quarks have to be sufficiently light.

By the `pure-gluonic deconfinement', FFHM meant the completely deconfined phase in pure Yang-Mills theory, because they focused on the 't Hooft large-$N_{\rm c}$ limit. Therefore, they claimed that a new phase exists between the completely deconfined and completely confined phases. This is precisely the partially deconfined phase.
\subsection{Transition temperatures}
Refs.~\cite{Hanada:2023krw,Hanada:2023rlk} discussed partial deconfinement in SU(3) QCD and estimated transition temperatures by using lattice configurations for $N_{\rm f}=2+1$ QCD generated by WHOT-QCD collaboration~\cite{Umeda:2012er}, as summarized in Table~\ref{table:transition_temperature_estimate}. The transition temperatures estimated in FFHM (firstly around 150 MeV, and then around 285 MeV) are roughly consistent with Table~\ref{table:transition_temperature_estimate}.

\begin{table}[hbtp]
  \centering
  \begin{tabular}{|c|c|c|}
    \hline
    Lattice size  & Temperature & Phase\\
    \hline 
    $4\times 32^3$   &  697 MeV & \textcolor{red}{CD}\\
    $6\times 32^3$   &  464 MeV & \textcolor{red}{CD}\\
    $8\times 32^3$   &  348 MeV & \textcolor{orange}{PD} or \textcolor{red}{CD}\\
    $10\times 32^3$  &  279 MeV & \textcolor{orange}{PD} \\
    $12\times 32^3$  &  232 MeV & \textcolor{orange}{PD}\\
    $14\times 32^3$  &  199 MeV & \textcolor{orange}{PD}\\
    $16\times 32^3$  &  174 MeV & \textcolor{blue}{CC} or \textcolor{orange}{PD}\\  
    \hline
  \end{tabular}
    \caption{
    A table from refs.~\cite{Hanada:2023krw,Hanada:2023rlk}. Lattice configurations for $N_{\rm f}=2+1$ QCD generated by WHOT-QCD collaboration~\cite{Umeda:2012er} were used to locate transition temperatures. Simulations were performed at lattice spacing $a\simeq 0.07$ fm, with up and down quarks heavier than physical mass (i.e., $\frac{m_\pi}{m_\rho}\simeq 0.63$ is larger than the real-world value $\frac{m_\pi}{m_\rho}\simeq 0.18$) and approximately physical strange-quark mass. The third column is the estimate of the phase discussed in~\cite{Hanada:2023krw,Hanada:2023rlk}. 
    \textcolor{red}{CD}, \textcolor{orange}{PD}, and \textcolor{blue}{CC} denote 
    \textcolor{red}{Complete Deconfinement}, 
    \textcolor{orange}{Partial Deconfinement}, and \textcolor{blue}{Complete Confinement}. 
}\label{table:transition_temperature_estimate}
\end{table}
\subsection{Quark dominance in partially deconfined phase}
Partial deconfinement tells us that when gluon and quark sectors are partially deconfined, the contribution from quarks can be larger than that of the gluons concerning the deconfined sector. As depicted in Fig.~\ref{fig:partial_deconf_in_components}, the numbers of deconfined degrees of freedom in the gluon and quark sectors scale as $M_{\rm deconf}^2$ and $M_{\rm deconf}N_{\rm f}$, respectively. 
If $M_{\rm deconf}\lesssim N_{\rm f}$ the dominant contribution comes from the quark sector. 

\begin{figure}[htbp]
\begin{center} 
\includegraphics[width=10cm]{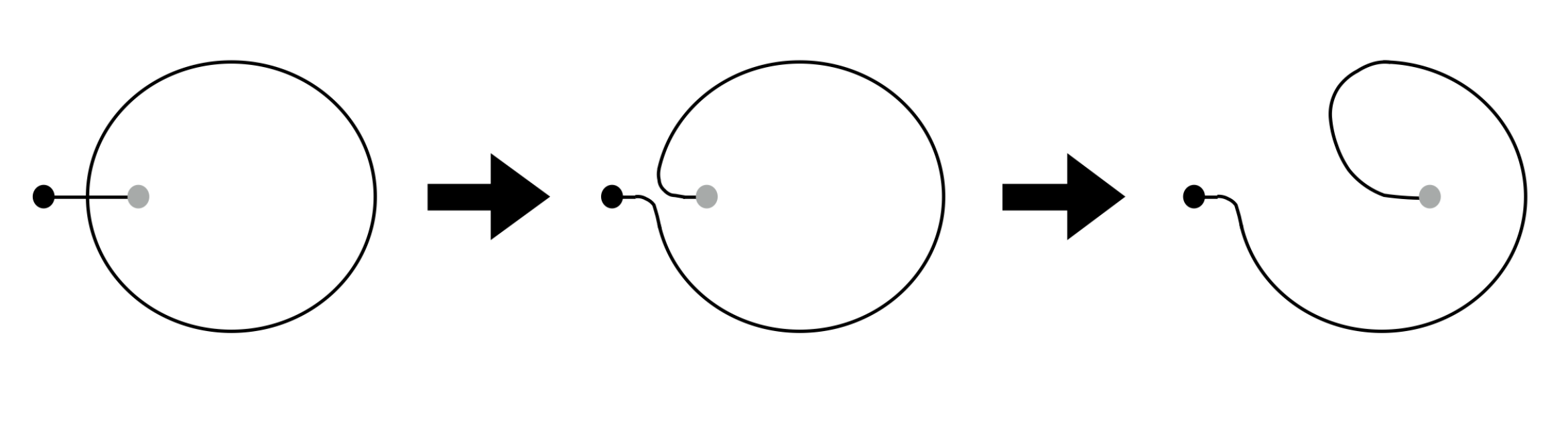}
\caption{
Deconfinement of a pair of probe quark and anti-quark in pure Yang-Mills theory. The condensation of many long closed strings characterizes deconfinement. Although the interaction at each intersection of strings is $1/N_{\rm c}$-suppressed, the interaction is not negligible in the deconfined phase because there are many intersections and hence small contributions from many intersections accumulate.  When a short open string that describes a pair of probe quark and anti-quark is introduced, the short open string interacts with a long closed string and forms a long open string. Once a long open string is formed, quark and anti-quark can easily be separated.  
In the same way, a long open string can interact with itself at the self-intersection points and form an open string and a closed string. Figure taken from ref.~\cite{Hanada:2014noa}. 
}\label{fig:probe_deconfinement}
\end{center}
\end{figure}

In contrast, FFHM claimed that only quarks deconfine. This is an unnatural scenario because, as soon as open strings (spaghetti) condense, closed strings appear through the interaction; see Fig.~\ref{fig:probe_deconfinement}.

FFHM claimed the SQGB phase disappears in the 't Hooft large-$N_{\rm c}$ limit ($N_{\rm f}$ fixed). This is the same as the partially deconfined phase. Partial deconfinement provides us with a more refined picture: this phase becomes the unstable saddle separating the completely deconfined phase and the completely confined phase. 

\subsection{Glueballs}
In ref.~\cite{Gautam:2022exf}, the strong-coupling limit of lattice gauge theory was studied, and the area law of the Polyakov loop correlator in the confined sector was demonstrated. 
Therefore, gluons in the confined sector form glueballs. When $\frac{M_{\rm deconf}}{N_{\rm c}}\ll 1$, most of the gluons are in the confined sector, and hence, most of the gluons form glueballs. 

FFHM claimed all gluons in the SQGB form glueballs, but partial deconfinement provides a more refined picture. Partial deconfinement is a valid picture up to $\frac{M_{\rm deconf}}{N_{\rm c}}=1$, and hence it can explain how the FFHM's picture can be improved so that the SQGB phase can be connected to the completely deconfined phase. 

\subsection{Chiral symmetry}
In the partially deconfined phase, a chiral condensate can be formed in the confined sector~\cite{Hanada:2019kue,Hanada:2021ksu}.\footnote{
To make the discussion in this section precise, the massless-quark limit is needed to have the exact chiral symmetry.
} As the deconfined sector grows ($M_{\rm deconf}$ goes up), the condensate melts and in the end completely disappears when $M_{\rm deconf}$ reaches $N_{\rm c}$. 
Chiral symmetry is broken in the partially deconfined phase in the sense that the condensate associated with the confined sector is not invariant under chiral transformation. However, a better way to view it is that \textit{the chiral symmetry is gradually restored through the partially deconfined phase, as the deconfined sector becomes larger}. In this sense, the CC$\to$PD point is the onset of chiral symmetry restoration.

FFHM claimed that chiral symmetry is restored in the SQGB phase. If we look at Fig.~9 in ref.~\cite{Fujimoto:2025sxx} (which was created based on ref.~\cite{Borsanyi:2010bp}), the value of the chiral condensate changes quickly in the SQGB phase. This is consistent with the situation described above --- the chiral symmetry is gradually restored through the partially deconfined phase --- by identifying the SQGB phase with the partially deconfined phase. 

\section{QCD in the Veneziano limit at finite baryon density}\label{sec:Veneziano-nonzero-mu}
As we have seen in Sec.~\ref{sec:Veneziano-zero-mu}, the proposed features of the SQGB phase fit well into the known features of the partially deconfined phase. A potentially new input to the partial deconfinement framework comes from the analysis of the finite chemical potential. FFHM speculated on the property of finite-$N_{\rm c}$ theory at finite temperature and finite baryon chemical potential, based on the 't Hooft large-$N_{\rm c}$ limit and quarkyonic phase~\cite{McLerran:2007qj}. This motivates us to study the properties of QCD at finite baryon chemical potential in the Veneziano limit. 

Note that, although our initial motivation came from the SQGB proposal, our analyses do not use anything from the SQGB proposal. 
\subsection{Baryon condensation as GWW transition}\label{sec:GWW-and-baryon-condensation}
Let the chemical potential per baryon be $\mu_{\rm B}=N_{\rm c}\mu_{\rm q}$ and the baryon mass be $m_{\rm B}=N_{\rm c}m$, where $\mu_{\rm q}$ and $m$ are $O(N_{\rm c}^0)$ quantities.\footnote{
Note that $m$ is defined as the baryon mass divided by $N_{\rm c}$, which differs from the quark mass in general. Although an analytical determination of \(m\) is challenging due to the strongly-coupled nature of the system, it is known to be of order $N_{\rm c}^0$~\cite{Witten:1979kh,Adkins:1983ya}.
} Then, the Boltzmann weight for each baryon is $\exp(N_{\rm c}\cdot\frac{\mu_{\rm q}-m}{T})$. At zero temperature, this Boltzmann weight is zero at $\mu_{\rm q}<m$ and infinite at $\mu_{\rm q}>m$. Therefore, baryons condense at $\mu_{\rm q}>m$.

As we saw in Sec.~\ref{sec:PD-review}, the constant offset in the Polyakov line phase distribution comes from a large symmetry in the extended Hilbert space, or equivalently, small gauge redundancy. Specifically, at vanishing chemical potential, a large stabilizer $\mathrm{SU}(N_{\rm c}-M_{\rm deconf})\subset\mathrm{SU}(N_{\rm c})$ for the SU($M_{\rm deconf}$)-deconfined states led to the offset. 

The situation is different for the baryon-condensed phase, because baryonic excitations form a nontrivial gauge orbit in the extended Hilbert space. When baryons condense, we do not find a particular stabilizer in the gauge group. This can be seen as follows. 
Firstly, let us see the symmetry of each baryon creation operator. Let the number of the $i$-th quark with spin $s=\uparrow$ or $\downarrow$ be $n_{i,s}$. By definition, $\sum_{i=1}^{N_{\rm f}}\sum_{s=\uparrow,\downarrow}n_{i,s}=N_{\rm c}$. For example, for $i=1=u$ (the up quark $u$) and spin $s=\uparrow$, we assign the color $1,2,\cdots,n_{u\uparrow}$ and act with $\hat{u}_{\uparrow,1}^\dagger\cdots\hat{u}_{\uparrow,n_{u\uparrow}}^\dagger$ on the ground state. This product is totally antisymmetric with respect to color indices (not over $1,2,\cdots,N_{\rm c}$, but over $1,2,\cdots,n_{u\uparrow}$) and hence invariant under SU($n_{u\uparrow}$). Taking into account all flavors and spins, each state has a stabilizer $\prod_{i=1}^{N_{\rm f}}\prod_{s=\uparrow,\downarrow}\mathrm{SU}(n_{i,s})$. When many baryons (specifically, density of order $N_{\rm c}$) condense, only the intersection of the stabilizers, which is negligibly small when $\eta$ is of order $N_{\rm c}^0$, is left as the genuine symmetry. 

Therefore, as in the complete deconfinement phase, the distribution of Polyakov line phases should be gapped when the baryons condense. Now we have the key observation in this paper: \emph{baryon condensation should be understood as the GWW transition}. Because baryons can condense without being associated with string condensation, the GWW transition can take place without string condensation, or equivalently, deconfinement.

The constant offset in the distribution of Polyakov line phases is the number of `active' degrees of freedom $M_{\rm active}$. Specifically, at finite baryon chemical potential, $M_{\rm active}$ is the number of degrees of freedom either in the deconfined sector or in the baryon condensation sector. By definition, 
\begin{align}
0\le M_{\rm deconf}\le M_{\rm active}\le N_{\rm c}\, .
\end{align}
In Fig.~\ref{fig:Mactive-vs-T-QCD} and Fig.~\ref{fig:Mdeconf-vs-T-QCD}, temperature dependence of $M_{\rm active}$ and $M_{\rm deconf}$ are depicted.

This provides us with a nontrivial feature:
\begin{itemize}
    \item 
    At a large chemical potential, $T_{\rm GWW}=T_{\rm B}$. The GWW transition is driven by baryon condensation. The transition is of first order, as we will see in Sec.~\ref{sec:QCD_critical_point}.

    \item 
    At zero chemical potential, $T_{\rm GWW}=T_{{\rm PD}\to{\rm CD}}$. The GWW transition is driven by string condensation. The transition is not of first order, as we have seen in Sec.~\ref{sec:Veneziano-zero-mu}.  		
\end{itemize}

With a mild assumption that there is a line of the GWW transition connecting these two regimes, we conclude that there is a \textit{critical point} where the first-order transition sets in. 
\subsection{Baryon condensation temperature}\label{sec:T_B-vs-mu_B}
To understand what happens at intermediate values of baryon chemical potential, let us see how the critical temperatures change with $\mu_{\rm q}$. 
Transition temperatures $T_{\rm PD\to CD}$ and $T_{\rm CC\to PD}$ should not be sensitive to $\mu_{\rm q}$, because they are driven by the condensation of open and closed strings, which are not directly coupled to $\mu_{\rm q}$. On the other hand, $T_{\rm B}$ can depend significantly on $\mu_{\rm q}$. Below, we estimate $T_{\rm B}$. 

In the Veneziano limit, because $N_{\rm f}$ is proportional to $N_{\rm c}$, there are many kinds of baryons. This gives a large entropy factor that competes with the Boltzmann weight $\exp(N_{\rm c}\cdot\frac{\mu_{\rm q}-m}{T})$~\cite{Hidaka:2008yy}. If the number of species of baryons times the Boltzmann weight exceeds 1, baryons condense. To compute this entropy factor, we count the number of species of baryons, which is the same as the number of ways $2N_{\rm f}$ flavor-spin pairs are assigned to $N_{\rm c}$ quarks, namely:
\begin{align}
    {N_{\rm c}+2N_{\rm f}-1 \choose 2N_{\rm f}-1}
    =
    \frac{(N_{\rm c}+2N_{\rm f}-1)!}{N_{\rm c}!\cdot(2N_{\rm f}-1)!}\, . 
\end{align}
Taking the logarithm, we obtain the `entropy':
\begin{align}
     \log\left(\frac{(N_{\rm c}+2N_{\rm f}-1)!}{N_{\rm c}!\cdot(2N_{\rm f}-1)!}\right)
     &\simeq
     (N_{\rm c}+2N_{\rm f})\log (N_{\rm c}+2N_{\rm f})
     -
     N_{\rm c}\log N_{\rm c}
     -
     2N_{\rm f}\log (2N_{\rm f})
     \nonumber\\
     &=
     \left\{(1+2\eta)\log(1+2\eta)-2\eta\log(2\eta)\right\}N_{\rm c}
     \nonumber\\
    &\equiv
     s_{\rm B}(\eta)\cdot N_{\rm c}\, . 
\label{eq:baryon_density_of_states_all}
\end{align}
Baryons can condense when the enhancement from this entropy contribution beats the suppression from the Boltzmann weight. Therefore, the baryons condense if
\begin{align}
    s_{\rm B}(\eta)
     +
     \frac{\mu_{\rm q}-m}{T}
     >
     0\, . 
\end{align}
The critical temperature is~\cite{Hidaka:2008yy}
\begin{align}
T_{\rm B}
=
\frac{m-\mu_{\rm q}}{s_{\rm B}(\eta)}
=
\frac{m-\mu_{\rm q}}{(1+2\eta)\log(1+2\eta)-2\eta\log(2\eta)}\, . 
\label{T_B-rough-estimate}
\end{align}
Note that the density of baryon below $T_{\rm B}$ is zero because $\left(s_{\rm B}(\eta)+\frac{\mu_{\rm q}-m}{T}\right)\cdot N_{\rm c}$ goes to negative infinity as $N_{\rm c}\to\infty$. Above $T_{\rm B}$, the energy density of the baryon condensate is of order $N_{\rm c}^2$ and hence the density of baryons is of order $N_{\rm c}^1$. Because there are $\sim e^{N_{\rm c}}$ kinds of baryons (see eq.~\eqref{eq:baryon_density_of_states_all}), the entropy density arising from the choice of baryons in the condensate is $N_{\rm c}^2$. Therefore, the transition at $T=T_{\rm B}$ is of first order. 

\begin{figure}[htbp]
	\centering
\includegraphics[width=10cm]{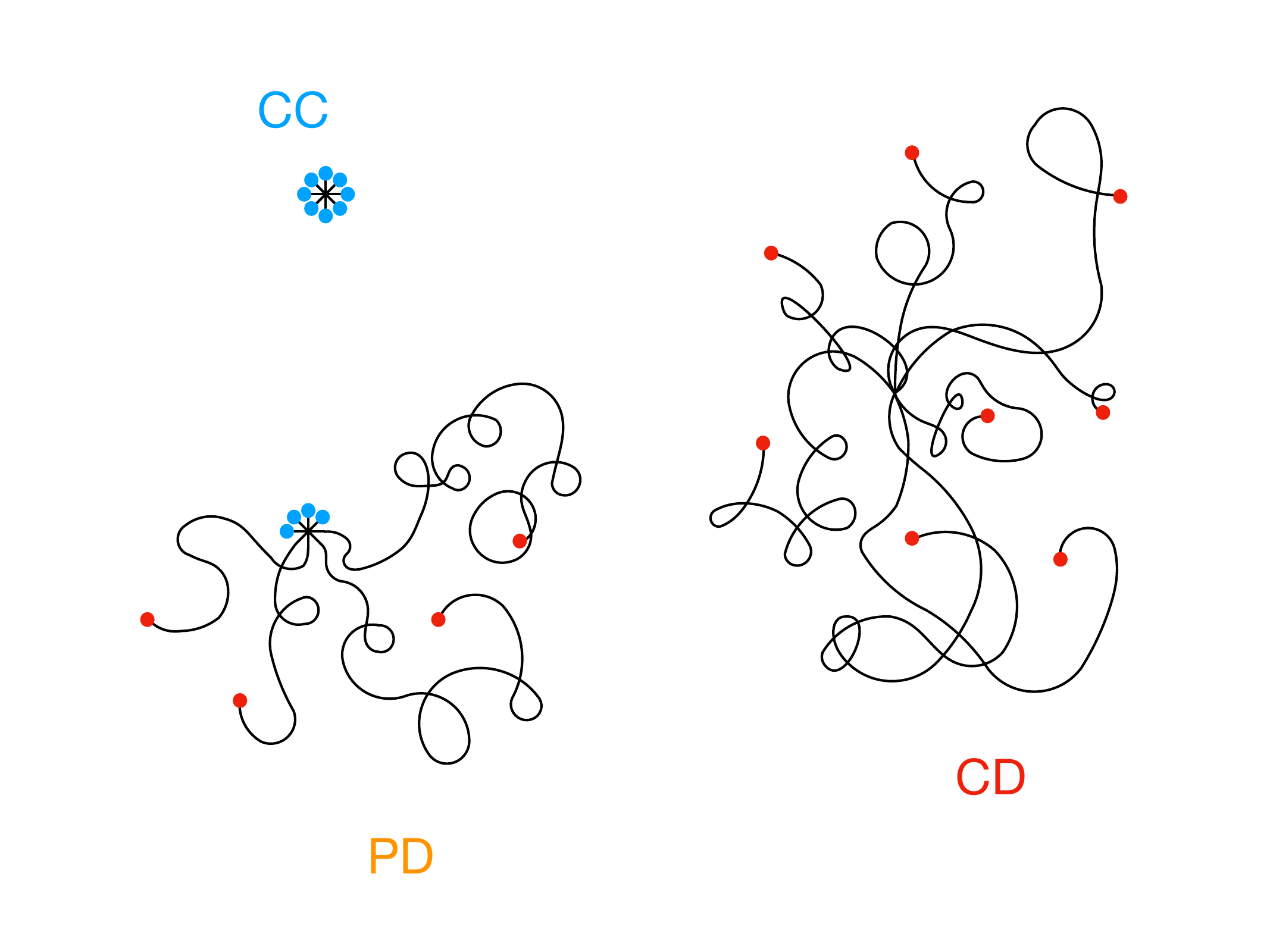}
	\caption{Speculative pictures of baryons in CC, PD, and CD phases. Blue and red points are quarks in the confined and deconfined sectors, respectively. If a baryon is excited in the partially or completely deconfined phase, it can interact with open or closed strings that are condensed and that constitute the deconfined sector. As a result, quarks in the deconfined sector can easily move around, while the quarks in the confined sector form a bound state.   }
	\label{fig:baryon}
\end{figure}

We should not take this estimate of $T_{\rm B}$ at face value, because there can be corrections associated with partial deconfinement. When some colors are deconfined and open and closed strings condense, long open strings can be attached to baryons (Fig.~\ref{fig:baryon}; see Sec.~\ref{sec:residual_symmetry_after_baryon_condensation} for a justification of this picture). The interaction between confined and deconfined sectors would alter $T_{\rm B}$, and the order of the phase transition may change. Furthermore, finite-$N_{\rm c}$ corrections can be significant. Nevertheless, we will use this expression as a crude estimate. To mimic the real world, we assume $m\sim 300$ -- $400$ MeV and $\eta\sim 0.5$ -- $1$, which gives $T_{\rm B}\sim 150$ -- $300$ MeV at $\mu_{\rm q}\ll m$. This temperature range is roughly the same as the partially deconfined regime. Therefore, qualitatively, we expect this transition to persist into the low-density region and merge with the PD$\to$CD transition at $\mu_{\rm q}=0$, forming a line of GWW transitions. This picture suggests interesting physics which we will discuss in Sec.~\ref{sec:QCD_critical_point}. 
(Note that we are taking the most `economical' scenario that does not introduce multiple new phase transitions.  In the case of 4D maximal super Yang-Mills at strong coupling, a very rich phase diagram with many phases has been predicted from the dual gravity analysis, and QCD may have a rich phase diagram as well.) 

\subsubsection*{Weak-coupling analysis}
Ref.~\cite{Hollowood:2011ep,Hollowood:2012nr} extended Schnitzer's analysis of weak-coupling QCD on small $\mathrm{S}^3$ in the Veneziano large-$N_\mathrm{c}$ limit to finite density. They studied the GWW transition using the saddle-point method.\footnote{
At $\mu_{\rm q}\neq 0$, the saddle becomes complex and the phase distribution at the saddle deviates from the unit circle. It is not clear whether the logic in ref.~\cite{Hanada:2020uvt} that relates the phase distribution to the symmetry of typical states in the extended Hilbert space is applicable to the complex saddle. Still, because the transition line is connected to the usual GWW transition at $\mu_{\rm q}\neq 0$, we can identify this transition with the usual GWW transition. 
}
They showed $T_{\rm GWW}$ behaving as we have proposed above, interpolating PD$\to$CD transition at $\mu_{\rm q}=0$ and baryon condensation at large $\mu_{\rm q}$; see Fig.~6 in ref.~\cite{Hollowood:2012nr}.\footnote{
Ref.~\cite{Hollowood:2011ep,Hollowood:2012nr} interpreted the GWW transition as the deconfinement transition which we disagree with. 
}
\subsection{Order of GWW phase transition and QCD critical point}\label{sec:QCD_critical_point}
The new phase in QCD was discussed in terms of partial deconfinement already in 2018~\cite{Hanada:2018zxn}, including the implication for the finite-density phase diagram and QCD critical point. Ref~\cite{Hanada:2018zxn} suggested that the emergence of the critical point can be understood as a change of deconfinement patterns depicted in Fig.~\ref{fig:three_patterns}, interpreting the GWW point as separating partial deconfinement from complete deconfinement. However, in this paper, we found that the meaning of the GWW point changes at a nonzero chemical potential. Therefore, the meaning of the red lines in Fig.~\ref{fig:three_patterns} should be reinterpreted in a more nuanced manner.

At large $\mu_{\rm q}$, the GWW temperature $T_{\rm GWW}$ is the baryon condensation temperature $T_{\rm B}$, which approaches zero. On the other hand, because mesons and glueballs are neutral in baryon charge, we expect that $T_{{\rm CC}\to{\rm PD}}$ and $T_{{\rm PD}\to{\rm CD}}$ are not so sensitive to the chemical potential. Therefore: 
\begin{align}
T_{\rm GWW} 
<
T_{{\rm CC}\to{\rm PD}}
<
T_{{\rm PD}\to{\rm CD}}
\qquad
\textrm{at\ large\ }\mu_{\rm q}\, . 
\end{align}
At $T_{\rm GWW}$, a first-order baryon-condensation transition takes place and the Polyakov line phase distribution jumps from an almost uniform distribution to a gapped distribution. This is the jump from CC-1 to CC-2 in the right panel of Fig.~\ref{fig:Mactive-vs-T-QCD}. At $T_{{\rm CC}\to{\rm PD}}$, partial deconfinement (pink line, PD-2) sets in, and complete deconfinement (red line, CD) is achieved at $T_{{\rm PD}\to{\rm CD}}$. 

At small nonzero $\mu_{\rm q}$, the GWW transition takes place in the partially deconfined phase: 
\begin{align}
T_{{\rm CC}\to{\rm PD}}
<
T_{\rm GWW} 
<
T_{{\rm PD}\to{\rm CD}}
\qquad
\textrm{at\ small\ }\mu_{\rm q}\, . 
\end{align}
The physics at $T_{\rm GWW}$ in this low-$\mu_{\rm q}$ region can be very different from that in the large-$\mu_{\rm q}$ region. Specifically, because the energy of order $N_{\rm c}^2$ is provided by partial deconfinement, a significant quantity of baryons can be excited already below $T_{\rm GWW}$, and the transition does not have to be of first order. Based on weak-coupling analysis and lattice simulations, we expect that the GWW transition is not of first order. The GWW transition splits the partially deconfined regime into two phases; see PD-1 and PD-2 in the left panel of Fig.~\ref{fig:Mactive-vs-T-QCD}. 

In between, a critical point appears where $T_{{\rm CC}\to{\rm PD}}$ and $T_{\rm GWW}$ cross each other. If we use \eqref{T_B-rough-estimate} as a crude estimate of $T_{\rm B}$, taking $T_{{\rm PD}\to{\rm CD}}\sim 350$ MeV and $T_{{\rm CC}\to{\rm PD}}\sim 175$ MeV based on Table~\ref{table:transition_temperature_estimate} (we ignore the $\mu_{\rm q}$ dependence of these temperatures) and $m\sim 310$ MeV, then the critical point is around $T_{\rm B}\sim T_{{\rm CC}\to{\rm PD}}\sim 175$ MeV and $\mu_{\rm q}\sim 155$MeV ($\mu_{\rm B}\sim 465$ MeV for $N_{\rm c}=3$). If $T_{{\rm CC}\to{\rm PD}}$ goes down slightly, say to 120 MeV, due to the back-reactions from baryons, and if we still assume \eqref{T_B-rough-estimate}, then $\mu_{\rm q}\sim 200$ MeV ($\mu_{\rm B}\sim 600$ MeV for $N_{\rm c}=3$). 

Note that the properties and even the existence of a QCD critical point have been rather controversial, and there is no consensus among researchers. This becomes clear if we compare QCD phase diagrams in the literature, e.g., Fig.~1 in a recent review~\cite{Du:2024wjm} and FFHM's proposal~\cite{Fujimoto:2025sxx} (Fig.~\ref{fig:FFHM_phase_diagram} in this paper). Although ref.~\cite{Du:2024wjm} showed the QCD critical point sitting close to our crude estimate, ref.~\cite{Fujimoto:2025sxx} did not show the QCD critical point.\footnote{The critical point in Fig.~\ref{fig:FFHM_phase_diagram} is a different one associated with the hadron liquid-gas transition.} Regarding the interpretation of the phases, ref.~\cite{Fujimoto:2025sxx} is closer to us.

\subsection{Survival of partial deconfinement above GWW}\label{sec:residual_symmetry_after_baryon_condensation}
At zero chemical potential, partial deconfinement was explained as a consequence of the residual `genuine' gauge symmetry~\cite{Hanada:2020uvt} (Sec.~\ref{sec:underlying_mechanism}). The SU($M_{\rm deconf}$)-deconfined states have unbroken SU($N_{\rm c}-M_{\rm deconf}$) `genuine' gauge symmetry as a stabilizer in the extended Hilbert space picture, and the constant offset in the Polyakov line phase distribution was proportional to $N_{\rm c}-M_{\rm deconf}$. Therefore, partial deconfinement was realized \textit{below} the GWW point. 

A natural question is: at finite density, can partial deconfinement --- in the sense of the block structure in Fig.~\ref{fig:partial_deconf_in_components} --- survive above the GWW point? We believe that the answer is yes, for the following reasons.

In principle, we can understand partial deconfinement by counting only gauge-invariant operators. In the simplest case of the Gaussian matrix model, to get states with energy $L$ up to ground-state energy, gauge-invariant operators consisting of products of $L$ matrices (Wilson loops with total length $L$) act on the Fock vacuum. The analyses in refs.~\cite{Hanada:2019czd,Berenstein:2018lrm} show that, up to the SU($N_{\rm c}$)-symmetrization, most of such states can be constructed using only the SU($M_{\rm deconf}$) subsector, where $L=M_{\rm deconf}^2/4$ (see Sec.~\ref{sec:underlying_mechanism}).\footnote{Recent developments regarding the trace relations at finite $N_{\rm c}$~\cite{deMelloKoch:2025ngs,deMelloKoch:2025qeq,deMelloKoch:2025rkw} would improve the understanding of such operators.} This property could be related to the constant offset in the Polyakov line phase distribution because the Fock vacuum was genuinely SU($N_{\rm c}$) symmetric. The same applies to QCD at zero chemical potential; the genuine SU($N_{\rm c}$) invariance of the ground state is the key to relate Polyakov line phases and partial deconfinement.  

Let us consider the large-$\mu_{\rm q}$ region. We start with CC-2 in  Fig.~\ref{fig:revised_picture_mu_vs_T} and see how deconfinement sets in as the temperature increases. To avoid double-counting, we take the SU($N_{\rm c}$)-symmetrized states at the low-temperature end of CC-2, which could be obtained, in principle, by acting with baryon creation operators on a genuinely SU($N_{\rm c}$) symmetric state and taking a complicated linear combination. This state is not genuinely SU($N_{\rm c}$) symmetric. To this state, we add stringy excitations to get excited states. Such states in CC-2 are not genuinely SU($N_{\rm c}$) symmetric either, and hence, the Polyakov line phase distribution has a gap. Still, when stringy excitations are concerned, they can be localized on the SU($M_{\rm deconf}$) sub-sector if the excitation is not too large; this is the property of string-creating operators acting on the baryon-condensed state, as discussed in Sec.~\ref{sec:underlying_mechanism}. 

A precise understanding of this phenomenon occurring at the GWW transition would require complicated computations of the dynamics. Intuitively, we could imagine a condensation of baryons with long strings as depicted in Fig.~\ref{fig:baryon}. 

\subsection{CC-2, PD-2, and quarkyonic phase}\label{sec:quarkyonic}
Because CC-2 and PD-2 in  Fig.~\ref{fig:revised_picture_mu_vs_T} continue to $\mu_{\rm q}>m$, it is natural to expect similarity with the quarkyonic phase~\cite{McLerran:2007qj}. The quarkyonic phase was studied in the 't Hooft large-$N_\mathrm{c}$ limit, where the partially deconfined phase becomes invisible (more precisely, becomes an unstable saddle). Therefore, CC-2 is the counterpart of the quarkyonic phase. PD-2 can be seen as interpolating the quarkyonic phase and completely deconfined phase: gluons and baryons gradually melt in PD-2 while strings are excited. 

\section{Discussions}\label{sec:discussion}
The phase diagram at finite chemical potential that we propose is only qualitative, and the actual phase diagram might be even richer. For example, the GWW line may be a zigzag, or the GWW transition and PD$\to$ CD transition may coincide for a finite range in the chemical potential, $0\le\mu_{\rm q}\le\, ^\exists\tilde{\mu}_{\rm q}$. In particular, for real-world QCD with nondegenerate mass, we do not expect our approach here to capture all the details. That said, do the qualitative features of the phase structure in the Veneziano limit survive in SU(3) QCD? We are optimistic because there can be distinct signals that are well defined at finite $N_{\rm c}$~\cite{Hanada:2021ksu,Hanada:2023krw,Hanada:2023rlk}, e.g., instanton condensation, chiral symmetry breaking/restoration (for the massless quark limit), and the condensations of Polyakov loops in various representations. 

For a precise understanding of the phase diagram, weak-coupling analyses analogous to refs.~\cite{Sundborg:1999ue,Aharony:2003sx,Schnitzer:2004qt,Hanada:2019czd} would play important roles. It is somewhat mysterious why weak-coupling analyses have been providing useful intuitions to strongly coupled theories. Recent lattice simulations~\cite{Rindlisbacher:2025dqw} of pure Yang-Mills theory that observed small confined-deconfined interface tensions suggest that a matrix model~\cite{Dumitru:2012fw} based on the weak-coupling description, which predicts vanishing tension, provides a good approximation of the strongly-coupled dynamics.\footnote{We thank Rob Pisarski for explaining this observation to us.} 

Another promising and tractable setup arises from studying 4D maximally supersymmetric Yang-Mills theory at strong coupling via its dual gravitational description \cite{Maldacena_1999}. In this context, the geometry of a small ten-dimensional black hole is thought to correspond to the partially deconfined phase of the gauge theory. Heuristically, this can be understood by noting that multiple small black holes can be positioned arbitrarily within the $\mathrm{AdS}_5\times\mathrm{S}^5$ spacetime, suggesting they originate from a small number of D-branes. However, such a multi-black hole configuration becomes inconsistent in the presence of a large AdS black hole, unless the small black holes are situated near the AdS boundary, which corresponds to a Higgsed phase in the gauge theory. Therefore, the large AdS black hole is naturally interpreted as the dual of the completely deconfined phase.

\subsubsection*{Potential implications for the superconformal index and supersymmetric black holes}
The findings presented above, in particular the refined understanding of the relationship between the GWW transition and the partially deconfined phase, might resolve some confusion concerning the holographic dual of the partially deconfined phase and the superconformal index. The superconformal index evaluated on $\mathcal{N}=4$ SYM on $\mathrm{S}^3\times\mathrm{S}^1$ should allow us to connect field theory configurations to points along the dual black hole phase diagram at different chemical potentials \cite{Kinney_2007, R_melsberger_2006, choi2024largeadsblackholes, Cabo_Bizet_2019, Benini:2018ywd} \footnote{To preserve supersymmetry, the true temperature of the system must remain at zero}. A natural question is to identify the point on the black hole phase diagram that corresponds to the GWW transition.

It was proposed in \cite{Hanada:2016pwv,Berenstein:2018lrm,Hanada:2018zxn} that the partially deconfined phase is dual to small black holes in $\mathrm{AdS}_5\times\mathrm{S}^5$. In the black hole phase diagram, a cusp in the free energy connects the large and small black holes. Small black holes are defined as lying below this cusp. BPS black holes follow a similar phase diagram, with a cusp in the phase diagram connecting large and small black holes \cite{Ezroura_2022}.

Previous work led to the strict identification of the GWW transition with the partially deconfined phase. Consequently, one would expect to locate the GWW transition at or close to this cusp.\footnote{There are two natural candidates for the PD/CD transition point: the border between large and small AdS$_5$ black holes wrapped on S$^5$, which is the cusp, and the Gregory-Laflamme transition between the small AdS$_5$ black hole and 10d black hole localized on S$^5$, which sits slightly below the cusp. 
} 
However, as was suggested in \cite{choi2024yang}, and which will be shown in more detail in upcoming work~\cite{Holden:2025}, computations of the superconformal index show that there is no GWW transition near the cusp. Moreover, the GWW transition is more likely located at energies of order $N_\mathrm{c}^0$. This appears starkly inconsistent with the physical understanding of the partially deconfined phase and its holographic duals.

Based on the reasoning presented in this paper, however, the GWW transition need not be identified with the onset of partial deconfinement. If objects condense that have nontrivial gauge orbits, just like baryons in QCD, the GWW transition will occur at lower energies or temperatures than expected, and the partially deconfined phase can continue to exist above the GWW transition. This could resolve the inconsistency. Possible candidates for the analogues of the baryons in $N=4$ SYM are fuzzy spheres and giant gravitons \cite{Myers_1999,McGreevy_2000,Grisaru:2000zn}.

As will be shown in the forthcoming work~\cite{Holden:2025}, a better characterization of the partially deconfined phase in this setting could be instanton condensation. Computations of the superconformal index show that instanton condensation does indeed occur in a region close to the cusp, consistent with expectations for the partially deconfined phase. Tunneling eigenvalues have been connected to giant gravitons in other regions of the phase diagram, perhaps offering a hint of evidence at the proposed resolution \cite{chen2024giantgravitonexpansioneigenvalue,eniceicu2024complexeigenvalueinstantonsfredholm}.

\section*{Acknowledgments}
\hspace{0.51cm}
The authors thank Yoshimasa Hidaka, Niko Jokela, Tatsuhiro Misumi, Rob Pisarski, Andreas Sch\"{a}fer, and Naoki Yamamoto for discussions and comments. 
M.~H.~thanks the STFC for the support through the consolidated grant ST/Z001072/1.
The work of H.W. was partially supported by Japan Society for the Promotion of Science (JSPS) KAKENHI Grant No. 24K00630.

\appendix
\section{Notation Summary}\label{sec:notation}

This subsection summarizes the key notation and symbols used throughout this paper.

\subsubsection*{Phase Labels}
\begin{itemize}
\item CC: Complete Confinement / Completely Confined phase
\item PD: Partial Deconfinement / Partially Deconfined phase 
\item CD: Complete Deconfinement / Completely Deconfined phase 
\item PD-1: Partially Deconfined phase 1 (below GWW transition)
\item PD-2: Partially Deconfined phase 2 (above GWW transition)
\item CC-1: Completely Confined phase 1 (below GWW transition, no baryon condensation)
\item CC-2: Completely Confined phase 2 (above GWW transition, with baryon condensation)
\end{itemize}

\subsubsection*{Transition Temperatures}
\begin{itemize}
\item $T_{\text{CC}\to\text{PD}}$: Transition temperature from Complete Confinement to Partial Deconfinement
\item $T_{\text{PD}\to\text{CD}}$: Transition temperature from Partial Deconfinement to Complete Deconfinement
\item $T_{\text{GWW}}$: Gross-Witten-Wadia transition temperature (gap opening in the distribution of Polyakov line phases)
\item $T_B$: Baryon condensation temperature
\end{itemize}

\subsubsection*{Physical Parameters}
\begin{itemize}
\item $T$: Temperature
\item $\mu_{\rm q}$: Quark chemical potential
\item $\mu_{\rm B}$: Baryon chemical potential, $\mu_{\rm B} = N_{\rm c} \mu_{\rm q}$
\item $m_{\rm B}$: Baryon mass
\item $m$: Baryon mass per quark, i.e., $m=m_{\rm B}/N_{\rm c}$. This is different from the quark mass. 
\item $N_{\rm c}$: Number of colors
\item $N_{\rm f}$: Number of flavors
\item $\eta$: Veneziano parameter, $\eta \equiv N_{\rm f}/N_{\rm c}$

\item $M_{\text{deconf}}$: Number of colors in the deconfined sector (string-condensed sector)
\item $M_{\rm active}$: Number of colors either in the deconfined sector (string-condensed sector) or baryon-condensed sector. 
\end{itemize}

\subsubsection*{Polyakov Loop and Eigenvalue Distribution}
\begin{itemize}
\item $\mathcal{P}$: Polyakov line, $\mathcal{P}=\mathrm{Path}\left(\exp\left(\mathrm{i}\int_0^\beta\mathrm{d}t A_t\right)\right)$, where $\mathrm{Path}$ stands for the path ordering.
\item $\theta$: Polyakov line phases, i.e., the eigenvalues of the Polyakov line $\mathcal{P}$ are $e^{\mathrm{i}\theta_j}$ ($j=1,2,\cdots,N_\mathrm{c}$), where $-\pi\le\theta_j<\pi$. 
\item $P$: Polyakov loop, $P \equiv \frac{1}{N_{\rm c}} \mathrm{Tr} \mathcal{P}= \frac{1}{N_{\rm c}}\sum_{j=1}^{N_{\rm c}} e^{\mathrm{i}\theta_j}$.
\item $\rho(\theta)$: The distribution of the Polyakov line phases normalized as $\int_{-\pi}^{+\pi}\mathrm{d}\theta\rho(\theta)=1$. 
\end{itemize}

\bibliographystyle{utphys}
\bibliography{ref}

@article{Witten:1979kh,
    author = "Witten, Edward",
    title = "{Baryons in the 1/n Expansion}",
    reportNumber = "HUTP-79-A007",
    doi = "10.1016/0550-3213(79)90232-3",
    journal = "Nucl. Phys. B",
    volume = "160",
    pages = "57--115",
    year = "1979"
}

@article{Adkins:1983ya,
    author = "Adkins, Gregory S. and Nappi, Chiara R. and Witten, Edward",
    title = "{Static Properties of Nucleons in the Skyrme Model}",
    reportNumber = "PRINT-83-0493 (IAS,PRINCETON)",
    doi = "10.1016/0550-3213(83)90559-X",
    journal = "Nucl. Phys. B",
    volume = "228",
    pages = "552",
    year = "1983"
}

@unpublished{Holden:2025,
  author = {Jack Holden},
  title  = {In preparation},
  year   = {2025}
}

@article{Hanada:2019rzv,
    author = "Hanada, Masanori and Ishiki, Goro and Watanabe, Hiromasa",
    title = "{Partial deconfinement in gauge theories}",
    eprint = "1911.11465",
    archivePrefix = "arXiv",
    primaryClass = "hep-lat",
    reportNumber = "UTHEP-739",
    doi = "10.22323/1.363.0055",
    journal = "PoS",
    volume = "LATTICE2019",
    pages = "055",
    year = "2019"
}

@article{Rindlisbacher:2025dqw,
    author = "Rindlisbacher, Tobias and Rummukainen, Kari and Salami, Ahmed",
  title = "{Confined-deconfined interface tension and latent heat in SU($N$) gauge theory}",
    eprint = "2506.15509",
    archivePrefix = "arXiv",
    primaryClass = "hep-lat",
    reportNumber = "HIP-2025-21/TH",
    month = "6",
    year = "2025"
}

@article{Dumitru:2012fw,
    author = "Dumitru, Adrian and Guo, Yun and Hidaka, Yoshimasa and Altes, Christiaan P. Korthals and Pisarski, Robert D.",
    title = "{Effective Matrix Model for Deconfinement in Pure Gauge Theories}",
    eprint = "1205.0137",
    archivePrefix = "arXiv",
    primaryClass = "hep-ph",
    reportNumber = "BNL-96946-2012-JA, NIKHEF-2012-002, RBRC-943, RIKEN-MP-40, RIKEN-QHP-20",
    doi = "10.1103/PhysRevD.86.105017",
    journal = "Phys. Rev. D",
    volume = "86",
    pages = "105017",
    year = "2012"
}

@article{Hanada:2021ksu,
    author = "Hanada, Masanori and Holden, Jack and Knaggs, Matthew and O'Bannon, Andy",
    title = "{Global symmetries and partial confinement}",
    eprint = "2112.11398",
    archivePrefix = "arXiv",
    primaryClass = "hep-th",
    reportNumber = "DMUS-MP-21/11",
    doi = "10.1007/JHEP03(2022)118",
    journal = "JHEP",
    volume = "03",
    pages = "118",
    year = "2022"
}

@article{Hanada:2019kue,
    author = "Hanada, Masanori and Robinson, Brandon",
    title = "{Partial-Symmetry-Breaking Phase Transitions}",
    eprint = "1911.06223",
    archivePrefix = "arXiv",
    primaryClass = "hep-th",
    doi = "10.1103/PhysRevD.102.096013",
    journal = "Phys. Rev. D",
    volume = "102",
    number = "9",
    pages = "096013",
    year = "2020"
}

@article{Schnitzer:2004qt,
    author = "Schnitzer, Howard J.",
    title = "{Confinement/deconfinement transition of large N gauge theories with N(f) fundamentals: N(f)/N finite}",
    eprint = "hep-th/0402219",
    archivePrefix = "arXiv",
    reportNumber = "BRX-TH-534",
    doi = "10.1016/j.nuclphysb.2004.06.057",
    journal = "Nucl. Phys. B",
    volume = "695",
    pages = "267--282",
    year = "2004"
}

@article{Gautam:2022exf,
    author = "Gautam, Vaibhav and Hanada, Masanori and Holden, Jack and Rinaldi, Enrico",
    title = "{Linear confinement in the partially-deconfined phase}",
    eprint = "2208.14402",
    archivePrefix = "arXiv",
    primaryClass = "hep-th",
    reportNumber = "DMUS-MP-22/14, RIKEN-iTHEMS-Report-22",
    doi = "10.1007/JHEP03(2023)195",
    journal = "JHEP",
    volume = "03",
    pages = "195",
    year = "2023"
}

@article{Fujimoto:2025sxx,
    author = "Fujimoto, Yuki and Fukushima, Kenji and Hidaka, Yoshimasa and McLerran, Larry",
    title = "{A New State of Matter between the Hadronic Phase and the Quark-Gluon Plasma?}",
    eprint = "2506.00237",
    archivePrefix = "arXiv",
    primaryClass = "hep-ph",
    reportNumber = "N3AS-25-009, RIKEN-iTHEMS-Report-25, YITP-25-80",
    month = "5",
    year = "2025"
}

@article{Hanada:2018zxn,
    author = "Hanada, Masanori and Ishiki, Goro and Watanabe, Hiromasa",
    title = "{Partial Deconfinement}",
    eprint = "1812.05494",
    archivePrefix = "arXiv",
    primaryClass = "hep-th",
    reportNumber = "UTHEP-728",
    doi = "10.1007/JHEP03(2019)145",
    journal = "JHEP",
    volume = "03",
    pages = "145",
    year = "2019",
    note = "[Erratum: JHEP 10, 029 (2019)]"
}

@article{Hanada:2016pwv,
    author = "Hanada, Masanori and Maltz, Jonathan",
    title = "{A proposal of the gauge theory description of the small Schwarzschild black hole in AdS$_5\times$S$^5$}",
    eprint = "1608.03276",
    archivePrefix = "arXiv",
    primaryClass = "hep-th",
    reportNumber = "SU-ITP-16-14, YITP-16-95",
    doi = "10.1007/JHEP02(2017)012",
    journal = "JHEP",
    volume = "02",
    pages = "012",
    year = "2017"
}

@article{Berenstein:2018lrm,
    author = "Berenstein, David",
    title = "{Submatrix deconfinement and small black holes in AdS}",
    eprint = "1806.05729",
    archivePrefix = "arXiv",
    primaryClass = "hep-th",
    doi = "10.1007/JHEP09(2018)054",
    journal = "JHEP",
    volume = "09",
    pages = "054",
    year = "2018"
}

@article{Hanada:2019czd,
    author = "Hanada, Masanori and Jevicki, Antal and Peng, Cheng and Wintergerst, Nico",
    title = "{Anatomy of Deconfinement}",
    eprint = "1909.09118",
    archivePrefix = "arXiv",
    primaryClass = "hep-th",
    doi = "10.1007/JHEP12(2019)167",
    journal = "JHEP",
    volume = "12",
    pages = "167",
    year = "2019"
}

@article{Hanada:2020uvt,
    author = "Hanada, Masanori and Shimada, Hidehiko and Wintergerst, Nico",
    title = "{Color confinement and Bose-Einstein condensation}",
    eprint = "2001.10459",
    archivePrefix = "arXiv",
    primaryClass = "hep-th",
    doi = "10.1007/JHEP08(2021)039",
    journal = "JHEP",
    volume = "08",
    pages = "039",
    year = "2021"
}

@article{Hanada:2023krw,
    author = "Hanada, Masanori and Ohata, Hiroki and Shimada, Hidehiko and Watanabe, Hiromasa",
    title = "{A New Perspective on Thermal Transition in QCD}",
    eprint = "2310.01940",
    archivePrefix = "arXiv",
    primaryClass = "hep-th",
    reportNumber = "YITP-23-121",
    doi = "10.1093/ptep/ptae044",
    journal = "PTEP",
    volume = "2024",
    number = "4",
    pages = "041B02",
    year = "2024"
}

@article{Hanada:2023rlk,
    author = "Hanada, Masanori and Watanabe, Hiromasa",
    title = "{On Thermal Transition in QCD}",
    eprint = "2310.07533",
    archivePrefix = "arXiv",
    primaryClass = "hep-th",
    reportNumber = "YITP-23-125",
    doi = "10.1093/ptep/ptae033",
    journal = "PTEP",
    volume = "2024",
    number = "4",
    pages = "043B02",
    year = "2024"
}

@article{Hanada:2014noa,
    author = "Hanada, Masanori and Maltz, Jonathan and Susskind, Leonard",
    title = "{Deconfinement transition as black hole formation by the condensation of QCD strings}",
    eprint = "1405.1732",
    archivePrefix = "arXiv",
    primaryClass = "hep-th",
    reportNumber = "YITP-14-35, SU-ITP-14-10, IPMU14-0106",
    doi = "10.1103/PhysRevD.90.105019",
    journal = "Phys. Rev. D",
    volume = "90",
    number = "10",
    pages = "105019",
    year = "2014"
}

@article{Hollowood:2011ep,
    author = "Hollowood, Timothy J. and Kumar, S. Prem and Myers, Joyce C.",
    title = "{Weak coupling large-N transitions at finite baryon density}",
    eprint = "1110.0696",
    archivePrefix = "arXiv",
    primaryClass = "hep-th",
    doi = "10.1007/JHEP11(2011)138",
    journal = "JHEP",
    volume = "11",
    pages = "138",
    year = "2011"
}

@article{McLerran:2007qj,
    author = "McLerran, Larry and Pisarski, Robert D.",
    title = "{Phases of cold, dense quarks at large N(c)}",
    eprint = "0706.2191",
    archivePrefix = "arXiv",
    primaryClass = "hep-ph",
    doi = "10.1016/j.nuclphysa.2007.08.013",
    journal = "Nucl. Phys. A",
    volume = "796",
    pages = "83--100",
    year = "2007"
}

@article{Hollowood:2012nr,
    author = "Hollowood, Timothy J. and Myers, Joyce C.",
    title = "{Deconfinement transitions of large N QCD with chemical potential at weak and strong coupling}",
    eprint = "1207.4605",
    archivePrefix = "arXiv",
    primaryClass = "hep-th",
    doi = "10.1007/JHEP10(2012)067",
    journal = "JHEP",
    volume = "10",
    pages = "067",
    year = "2012"
}

@article{Aharony:2003sx,
    author = "Aharony, Ofer and Marsano, Joseph and Minwalla, Shiraz and Papadodimas, Kyriakos and Van Raamsdonk, Mark",
    editor = "Doebner, H. D. and Dobrev, V. K.",
    title = "{The Hagedorn - deconfinement phase transition in weakly coupled large N gauge theories}",
    eprint = "hep-th/0310285",
    archivePrefix = "arXiv",
    reportNumber = "WIS-29-03-DPP",
    doi = "10.4310/ATMP.2004.v8.n4.a1",
    journal = "Adv. Theor. Math. Phys.",
    volume = "8",
    pages = "603--696",
    year = "2004"
}

@article{Borsanyi:2010bp,
    author = "Borsanyi, Szabolcs and Fodor, Zoltan and Hoelbling, Christian and Katz, Sandor D and Krieg, Stefan and Ratti, Claudia and Szabo, Kalman K.",
    collaboration = "Wuppertal-Budapest",
    title = "{Is there still any $T_c$ mystery in lattice QCD? Results with physical masses in the continuum limit III}",
    eprint = "1005.3508",
    archivePrefix = "arXiv",
    primaryClass = "hep-lat",
    reportNumber = "WUB-10-11, MIT-CTP-4152",
    doi = "10.1007/JHEP09(2010)073",
    journal = "JHEP",
    volume = "09",
    pages = "073",
    year = "2010"
}

@article{Witten:1998zw,
    author = "Witten, Edward",
    editor = "Bergstrom, L. and Lindstrom, U.",
    title = "{Anti-de Sitter space, thermal phase transition, and confinement in gauge theories}",
    eprint = "hep-th/9803131",
    archivePrefix = "arXiv",
    reportNumber = "IASSNS-HEP-98-21",
    doi = "10.4310/ATMP.1998.v2.n3.a3",
    journal = "Adv. Theor. Math. Phys.",
    volume = "2",
    pages = "505--532",
    year = "1998"
}

@article{Aharony:1999ti,
    author = "Aharony, Ofer and Gubser, Steven S. and Maldacena, Juan Martin and Ooguri, Hirosi and Oz, Yaron",
    title = "{Large N field theories, string theory and gravity}",
    eprint = "hep-th/9905111",
    archivePrefix = "arXiv",
    reportNumber = "CERN-TH-99-122, HUTP-99-A027, LBNL-43113, RU-99-18, UCB-PTH-99-16, LBL-43113",
    doi = "10.1016/S0370-1573(99)00083-6",
    journal = "Phys. Rept.",
    volume = "323",
    pages = "183--386",
    year = "2000"
}

@article{Sundborg:1999ue,
    author = "Sundborg, Bo",
    title = "{The Hagedorn transition, deconfinement and N=4 SYM theory}",
    eprint = "hep-th/9908001",
    archivePrefix = "arXiv",
    reportNumber = "USITP-99-06",
    doi = "10.1016/S0550-3213(00)00044-4",
    journal = "Nucl. Phys. B",
    volume = "573",
    pages = "349--363",
    year = "2000"
}

@article{Hidaka:2008yy,
    author = "Hidaka, Yoshimasa and McLerran, Larry D. and Pisarski, Robert D.",
    title = "{Baryons and the phase diagram for a large number of colors and flavors}",
    eprint = "0803.0279",
    archivePrefix = "arXiv",
    primaryClass = "hep-ph",
    doi = "10.1016/j.nuclphysa.2008.05.009",
    journal = "Nucl. Phys. A",
    volume = "808",
    pages = "117--123",
    year = "2008"
}

@article{Veneziano:1976wm,
    author = "Veneziano, G.",
    title = "{Some Aspects of a Unified Approach to Gauge, Dual and Gribov Theories}",
    reportNumber = "CERN-TH-2200",
    doi = "10.1016/0550-3213(76)90412-0",
    journal = "Nucl. Phys. B",
    volume = "117",
    pages = "519--545",
    year = "1976"
}

@article{Cohen:2023hbq,
    author = "Cohen, T. D. and Glozman, L. Ya.",
    title = "{Large $N_c$ QCD phase diagram at $\mu _B=0$}",
    eprint = "2311.07333",
    archivePrefix = "arXiv",
    primaryClass = "hep-ph",
    doi = "10.1140/epja/s10050-024-01400-9",
    journal = "Eur. Phys. J. A",
    volume = "60",
    number = "9",
    pages = "171",
    year = "2024"
}

@article{Glozman:2019fku,
    author = "Glozman, L. Ya",
    title = "{Three regimes of QCD}",
    eprint = "1907.01820",
    archivePrefix = "arXiv",
    primaryClass = "hep-ph",
    doi = "10.1142/S0217751X20440315",
    journal = "Int. J. Mod. Phys. A",
    volume = "36",
    number = "25",
    pages = "2044031",
    year = "2021"
}

@article{Umeda:2012er,
    author = "Umeda, T. and Aoki, S. and Ejiri, S. and Hatsuda, T. and Kanaya, K. and Maezawa, Y. and Ohno, H.",
    collaboration = "WHOT-QCD",
    title = "{Equation of state in 2+1 flavor QCD with improved Wilson quarks by the fixed scale approach}",
    eprint = "1202.4719",
    archivePrefix = "arXiv",
    primaryClass = "hep-lat",
    doi = "10.1103/PhysRevD.85.094508",
    journal = "Phys. Rev. D",
    volume = "85",
    pages = "094508",
    year = "2012"
}

@article{deMelloKoch:2025ngs,
    author = "de Mello Koch, Robert and Jevicki, Antal",
    title = "{Structure of loop space at finite N}",
    eprint = "2503.20097",
    archivePrefix = "arXiv",
    primaryClass = "hep-th",
    doi = "10.1007/JHEP06(2025)011",
    journal = "JHEP",
    volume = "06",
    pages = "011",
    year = "2025"
}

@article{deMelloKoch:2025qeq,
    author = "de Mello Koch, Robert and Kim, Minkyoo and Van Zyl, Hendrik J. R.",
    title = "{From Symmetry to Structure: Gauge-Invariant Operators in Multi-Matrix Quantum Mechanics}",
    eprint = "2507.01219",
    archivePrefix = "arXiv",
    primaryClass = "hep-th",
    month = "7",
    year = "2025"
}

@article{Du:2024wjm,
    author = "Du, Lipei and Sorensen, Agnieszka and Stephanov, Mikhail",
    title = "{The QCD phase diagram and Beam Energy Scan physics: A theory overview}",
    eprint = "2402.10183",
    archivePrefix = "arXiv",
    primaryClass = "nucl-th",
    reportNumber = "INT-PUB-24-017",
    doi = "10.1142/9789811294679_0007",
    journal = "Int. J. Mod. Phys. E",
    volume = "33",
    number = "07",
    pages = "2430008",
    year = "2024"
}

@article{Nagata:2021ugx,
    author = "Nagata, Keitaro",
    title = "{Finite-density lattice QCD and sign problem: Current status and open problems}",
    eprint = "2108.12423",
    archivePrefix = "arXiv",
    primaryClass = "hep-lat",
    doi = "10.1016/j.ppnp.2022.103991",
    journal = "Prog. Part. Nucl. Phys.",
    volume = "127",
    pages = "103991",
    year = "2022"
}

@article{deForcrand:2009zkb,
    author = "de Forcrand, Philippe",
    editor = "Liu, Chuan and Zhu, Yu",
    title = "{Simulating QCD at finite density}",
    eprint = "1005.0539",
    archivePrefix = "arXiv",
    primaryClass = "hep-lat",
    reportNumber = "CERN-PH-TH-2010-090",
    doi = "10.22323/1.091.0010",
    journal = "PoS",
    volume = "LAT2009",
    pages = "010",
    year = "2009"
}

@article{R_melsberger_2006,
   title={Counting chiral primaries in , superconformal field theories},
   volume={747},
   ISSN={0550-3213},
   url={http://dx.doi.org/10.1016/j.nuclphysb.2006.03.037},
   DOI={10.1016/j.nuclphysb.2006.03.037},
   number={3},
   journal={Nuclear Physics B},
   publisher={Elsevier BV},
   author={Römelsberger, Christian},
   year={2006},
   month=jul, pages={329–353} 
}

@article{Kinney_2007,
   title={An Index for 4 Dimensional Super Conformal Theories},
   volume={275},
   ISSN={1432-0916},
   url={http://dx.doi.org/10.1007/s00220-007-0258-7},
   DOI={10.1007/s00220-007-0258-7},
   number={1},
   journal={Communications in Mathematical Physics},
   publisher={Springer Science and Business Media LLC},
   author={Kinney, Justin and Maldacena, Juan and Minwalla, Shiraz and Raju, Suvrat},
   year={2007},
   month=jun, pages={209–254} 
}

@article{Ezroura_2022,
   title={The phase diagram of BPS black holes in AdS5},
   volume={2022},
   ISSN={1029-8479},
   url={http://dx.doi.org/10.1007/JHEP09(2022)033},
   DOI={10.1007/jhep09(2022)033},
   number={9},
   journal={Journal of High Energy Physics},
   publisher={Springer Science and Business Media LLC},
   author={Ezroura, Nizar and Larsen, Finn and Liu, Zhihan and Zeng, Yangwenxiao},
   year={2022},
   month=sep 
}

@article{choi2024yang,
  title={The Yang-Mills duals of small AdS black holes},
  author={Choi, Sunjin and Jeong, Saebyeok and Kim, Seok},
  journal={Journal of High Energy Physics},
  volume={2024},
  number={7},
  pages={1--60},
  year={2024},
  publisher={Springer}
}

@misc{chen2024giantgravitonexpansioneigenvalue,
      title={Giant graviton expansion from eigenvalue instantons}, 
      author={Yiming Chen and Raghu Mahajan and Haifeng Tang},
      year={2024},
      eprint={2407.08155},
      archivePrefix={arXiv},
      primaryClass={hep-th},
      url={https://arxiv.org/abs/2407.08155}, 
}

@misc{eniceicu2024complexeigenvalueinstantonsfredholm,
      title={Complex eigenvalue instantons and the Fredholm determinant expansion in the Gross-Witten-Wadia model}, 
      author={Dan Stefan Eniceicu and Raghu Mahajan and Chitraang Murdia},
      year={2024},
      eprint={2308.06320},
      archivePrefix={arXiv},
      primaryClass={hep-th},
      url={https://arxiv.org/abs/2308.06320}, 
}

@article{Cabo_Bizet_2019,
   title={Microscopic origin of the Bekenstein-Hawking entropy of supersymmetric AdS5 black holes},
   volume={2019},
   ISSN={1029-8479},
   url={http://dx.doi.org/10.1007/JHEP10(2019)062},
   DOI={10.1007/jhep10(2019)062},
   number={10},
   journal={Journal of High Energy Physics},
   publisher={Springer Science and Business Media LLC},
   author={Cabo-Bizet, Alejandro and Cassani, Davide and Martelli, Dario and Murthy, Sameer},
   year={2019},
   month=oct }

@misc{choi2024largeadsblackholes,
      title={Large AdS black holes from QFT}, 
      author={Sunjin Choi and Joonho Kim and Seok Kim and June Nahmgoong},
      year={2024},
      eprint={1810.12067},
      archivePrefix={arXiv},
      primaryClass={hep-th},
      url={https://arxiv.org/abs/1810.12067}, 
}

@article{Benini:2018ywd,
    author = "Benini, Francesco and Milan, Elisa",
    title = "{Black Holes in 4D $\mathcal{N}$=4 Super-Yang-Mills Field Theory}",
    eprint = "1812.09613",
    archivePrefix = "arXiv",
    primaryClass = "hep-th",
    reportNumber = "SISSA 56/2018/FISI",
    doi = "10.1103/PhysRevX.10.021037",
    journal = "Phys. Rev. X",
    volume = "10",
    number = "2",
    pages = "021037",
    year = "2020"
}

@article{McGreevy_2000,
   title={Invasion of the giant gravitons from anti-de Sitter space},
   volume={2000},
   ISSN={1029-8479},
   url={http://dx.doi.org/10.1088/1126-6708/2000/06/008},
   DOI={10.1088/1126-6708/2000/06/008},
   number={06},
   journal={Journal of High Energy Physics},
   publisher={Springer Science and Business Media LLC},
   author={McGreevy, John and Susskind, Leonard and Toumbas, Nicolaos},
   year={2000},
   month=jun, pages={008–008} 
}

@article{Grisaru:2000zn,
    author = "Grisaru, Marcus T. and Myers, Robert C. and Tafjord, Oyvind",
    title = "{SUSY and goliath}",
    eprint = "hep-th/0008015",
    archivePrefix = "arXiv",
    reportNumber = "MCGILL-00-21, BRX-TH-472",
    doi = "10.1088/1126-6708/2000/08/040",
    journal = "JHEP",
    volume = "08",
    pages = "040",
    year = "2000"
}

@article{Myers_1999,
   title={Dielectric-branes},
   volume={1999},
   ISSN={1029-8479},
   url={http://dx.doi.org/10.1088/1126-6708/1999/12/022},
   DOI={10.1088/1126-6708/1999/12/022},
   number={12},
   journal={Journal of High Energy Physics},
   publisher={Springer Science and Business Media LLC},
   author={Myers, Robert C},
   year={1999},
   month=dec, pages={022–022} 
}

@article{Gross:1980he,
      author         = "Gross, D. J. and Witten, Edward",
      title          = "{Possible Third Order Phase Transition in the Large N
                        Lattice Gauge Theory}",
      journal        = "Phys. Rev.",
      volume         = "D21",
      year           = "1980",
      pages          = "446-453",
      doi            = "10.1103/PhysRevD.21.446",
      SLACcitation   = "%%CITATION = PHRVA,D21,446;%%"
}

@article{Wadia:2012fr,
      author         = "Wadia, Spenta R.",
      title          = "{A Study of U(N) Lattice Gauge Theory in 2-dimensions}",
      year           = "2012",
      eprint         = "1212.2906",
      archivePrefix  = "arXiv",
      primaryClass   = "hep-th",
      reportNumber   = "ICTS-2012-13, TIFR-TH-2012-47",
      SLACcitation   = "%%CITATION = ARXIV:1212.2906;%%"
}

@article{deMelloKoch:2025rkw,
    author = "de Mello Koch, Robert and Jevicki, Antal",
    title = "{Hilbert Space of Finite $N$ Multi-matrix Models}",
    eprint = "2508.11986",
    archivePrefix = "arXiv",
    primaryClass = "hep-th",
    month = "8",
    year = "2025"
}

@article{Maldacena_1999,
   title={The Large-N Limit of Superconformal Field Theories and Supergravity},
   volume={38},
   ISSN={1572-9575},
   url={http://dx.doi.org/10.1023/A:1026654312961},
   DOI={10.1023/a:1026654312961},
   number={4},
   journal={International Journal of Theoretical Physics},
   publisher={Springer Science and Business Media LLC},
   author={Maldacena, Juan},
   year={1999},
   month=apr, pages={1113–1133} }

\end{document}